\documentclass[nofootinbib,prd,showpacs]{revtex4-2}
\usepackage{amsmath}
\usepackage{amsfonts}
\usepackage{amssymb}
\usepackage{graphicx}
\graphicspath{ {Fig/} }
\usepackage{ulem}
\usepackage[usenames,dvipsnames]{xcolor}
\usepackage{hyperref}
\hypersetup{colorlinks=true,
citecolor=MidnightBlue, linkcolor=MidnightBlue, urlcolor=Cyan}
\usepackage{tikz}
\definecolor{purple}{rgb}{1,0,1}
\definecolor{lime}{HTML}{A6CE39} 

\def\be{\begin{equation}}
\def\ee{\end{equation}}
\def\beq{\begin{eqnarray}}
\def\eeq{\end{eqnarray}}

\newcommand{\orcidicon}{%
	\begin{tikzpicture}
	\draw[lime, fill=lime] (0,0)
		circle [radius=0.16]
		node[white] {{\fontfamily{qag}\selectfont \tiny ID}};
	\draw[white, fill=white] (-0.0625,0.095)
		circle [radius=0.007];
	\end{tikzpicture}	\hspace{-2mm}
}
\newcommand\orcidHilberto{{\href{https://orcid.org/0000-0003-1078-6118}{\orcidicon}}}
\newcommand\orcidTiberiu{{\href{https://orcid.org/0000-0002-1990-9172}{\orcidicon}}}
\newcommand\orcidRosa{{\href{https://orcid.org/0000-0003-4148-7372}{\orcidicon}}}\newcommand\orcidFrancisco{{\href{https://orcid.org/0000-0002-9388-8373}{\orcidicon}}}

\begin{document}

\title{Cosmic strings in generalized hybrid metric-Palatini gravity}

\author{Hilberto M. R. da Silva\orcidHilberto\!\!}
\email{hilberto.silva@astro.up.pt}
\affiliation{Instituto de Astrof\'{i}sica e Ci\^encias do Espa\c{c}o, Universidade do Porto, CAUP, Rua das Estrelas, PT4150-762 Porto, Portugal, Centro de Astrof\'{i}sica da Universidade do Porto,
Rua das Estrelas, PT4150-762 Porto, Portugal}
 \author{Tiberiu Harko\orcidTiberiu\!\!}
 \email{tiberiu.harko@aira.astro.ro}
 \affiliation{Department of Theoretical Physics, National Institute of Physics
and Nuclear Engineering (IFIN-HH), Bucharest, 077125 Romania,}
 \affiliation{Astronomical Observatory, 19 Ciresilor Street, 400487 Cluj-Napoca, Romania,}
\affiliation{Department of Physics, Babes-Bolyai University, Kogalniceanu Street,
Cluj-Napoca 400084, Romania}
\affiliation{School of Physics, Sun Yat-Sen University, Xingang Road, Guangzhou 510275, People's
Republic of China,}
\author{Francisco S. N. Lobo\orcidFrancisco\!\!}
\email{fslobo@fc.ul.pt}
\affiliation{Instituto de Astrof\'{i}sica e Ci\^{e}ncias do Espa\c{c}o, Faculdade de Ci\^encias da Universidade de Lisboa, Edif\'{i}cio C8, Campo Grande, P-1749-016, Lisbon, Portugal}
\affiliation{Departamento de F\'{i}sica, Faculdade de Ci\^{e}ncias, Universidade de Lisboa, Edifício C8, Campo Grande, PT1749-016 Lisbon, Portugal}
\author{Jo\~{a}o Lu\'{i}s Rosa\orcidRosa\!\!}
\email{joaoluis92@gmail.com}
\affiliation{Institute of Physics, University of Tartu, W. Ostwaldi 1, 50411 Tartu, Estonia}

\date{\today}

\begin{abstract}
We consider the possible existence of gravitationally bound stringlike objects, and approximation of local $U(1)$ Abelian-Higgs strings, in the framework of the generalized hybrid metric-Palatini gravity theory, in which the gravitational action is represented by an arbitrary function of the Ricci and of the Palatini scalars, respectively. The theory admits an equivalent scalar-tensor representation in terms of two independent scalar fields.  Assuming cylindrical symmetry, and the boost invariance of the metric, we obtain the gravitational field equations that describe cosmic stringlike structures in the theory. The physical and geometrical properties of the cosmic strings are determined by the two scalar fields, as well by an effective field potential, functionally dependent on both scalar fields. The field equations can be exactly solved for a vanishing, and a constant potential, respectively, with the corresponding string tension taking both negative and positive values. Furthermore, for more general classes of potentials, having an additive and a multiplicative algebraic structure in the two scalar fields, the gravitational field equations are solved numerically. For each potential we investigate the effects of the variations of the potential parameters and of the boundary conditions on the structure of the cosmic string. In this way, we obtain a large class of stable stringlike astrophysical configurations, whose basic parameters (string tension and radius) depend essentially on the effective field potential, and on the boundary conditions.
\end{abstract}


\maketitle
\tableofcontents

\section{Introduction}\label{Introduction}

The concept of symmetry is a very powerful one in Physics. The idea that a physical law remains invariant under groups of transformations has taken us very far \cite{Martin:2003cv}, as the theories of relativity can attest. Hand in hand with the concept of symmetry comes the concept of spontaneous symmetry breaking, in which a system, undergoing a phase transition, no longer retains the symmetry it showed in a higher energy state.

Spontaneous symmetry breaking is central to the idea of a Grand Unified Theory (GUT), in which the strong and electroweak interactions are unified and was inspired by the success of electroweak theory \cite{Weinberg:1967tq, Salam:1959zz,Salam:1964ry}, which unifies weak and electromagnetic interactions by the gauge group $SU(2) \times U(1)$ at a scale of $10^2 GeV$. One or more phase transitions may have occurred as the Universe expanded, originating the different groups of symmetries exhibited by the Standard Model today.
A GUT for Particle Physics is supported by the fact that the coupling ``constants'' of the Standard Model seems to vary slowly with energy and converge to a common value at higher energies, somewhere in the region of $10^{16}$GeV \cite{Amaldi:1991zx}, the Grand Unification Scale. In several of the Grand Unified Scenarios that have been proposed, a ``universal covering group'' $G$ would be effective above the GUT scale but its symmetry would spontaneously break into the Standard Model $SU(3)\times SU(2)\times U(1)$, where $SU(3)$ is the symmetry group of quantum chromodynamics, describing the strong interaction, and $SU(2)\times U(1)$ is the already mentioned electroweak group.
These phase transitions in the early Universe may give rise to a network of topological defects through the so-called Kibble mechanism \cite{Kibble:1976sj}.

To illustrate the Kibble mechanism that leads to the formation of topological defects, in particular, cosmic strings, let's consider a simple Goldstone model, with a single complex scalar field, $\phi$, with a classical Lagrangian density:

\begin{equation}
\mathcal{L}=-\partial_\mu \bar{\phi}\partial ^\mu \phi - V(\phi)\,,
\label{lagrangiangoldstone}
\end{equation}
where the potential is given by

\begin{equation}
V(\phi)=\frac{1}{4}\lambda\left(\bar{\phi}\phi-\eta^2\right)^2\,,
\end{equation}
with $\eta$ and $\lambda$ being positive constants. This type of potential is known as ``Mexican hat potential'' for the complex scalar field $\phi$. The Lagrangian \eqref{lagrangiangoldstone} is invariant under global $U(1)$ transformations.  It was shown \cite{Weinberg:1974hy,Kirzhnits:1974as,Dolan:1973qd} that this type of Lagrangian leads to an effective high temperature-dependent potential that can be described as
\begin{equation}
V_{eff}(\phi,T)=-\frac{\lambda}{12}\left(T^2-6\eta^2\right)|\phi|^2+\frac{\lambda}{4}|\phi|^4 \,,
\end{equation}

As $T$ approaches $T_c=\sqrt{6}\eta$ from above we have a VEV for the field $\langle |\phi|\rangle=0$, and the symmetry is restored, while for $T<T_c$,  the VEV for the scalar field is no longer zero and so the symmetry is spontaneously broken. With decreasing temperature, the field ``rolls'' to the new VEV, thermal and quantum fluctuations will dictate the new vacuum state and, as the minima are now degenerate, different patches of the Universe, separated by a distance larger than the particle horizon, $R_H=a(t)\int_{0}^{t} 1/a(t') \,dt'$, where $a\left(t\right)$ is a scale factor, in the time period of the symmetry breaking, may roll to different, but equivalent, vacua states.

As different patches start to grow through expansion, they eventually become causally connected, and in the limits of these patches the field transition is not smooth and a region of false vacuum becomes ``trapped'': a topological defect is formed. And so the Universe may contain several patches of true vacuum domains, in a sea of false-vacuum domains, a network of topological defects. For a more detailed description, we refer the reader to \cite{Vilenkin:2000jqa}.

The array of possible types of topological defects, the so-called topological defects zoo, is vast and will essentially depend on the topology of the Minima Manifold. Let us consider a field obeying a symmetry defined by a group $G$ (universal covering group) that is broken to a subgroup $K$, i.e., $G\rightarrow K$. In this case, we can classify the topological defects formed by looking at the $n$-th homotopy group $\pi_n$ of the quotient space $G/K$. For non-trivial $\pi_n(G/K)$, we can have domain walls for $n=0$, cosmic strings for $n=1$, monopoles for $n=2$, or textures for $n=3$.
While some of these defects tend to be inherently unstable \cite{Vilenkin:2000jqa}, domain walls and monopoles are cosmologically catastrophic or severely constrained by observations \cite{1991ApJ...373L..35N}. The case of monopoles is interesting since all GUTs based on simple gauge groups lead to the formation of topologically stable monopoles whose density is approximately $10^{18}$ times greater than the experimental limit \cite{Jeannerot:2003qv}, rendering the models unsustainable, except if the transition occurs before, or at the beginning of, inflation.

Hence, the cosmic inflationary scenario, being a crucial ingredient in modern cosmology, solving some important issues such as the flatness problem or the horizon problem, will also play an important role in the survival of a topological defect network, since the relation between the energy scale at which the SSB takes place and the one where inflation occurs will determine if either the network of topological defects becomes diluted, which is important in the case of domain walls or monopoles, or become energetically dominant, changing drastically the evolution of the Universe. However, a network of cosmic strings produced at the end of inflation, expected for a variety of GUT scenarios \cite{Jeannerot:2003qv}, should be stable and good candidates for further analysis.

That being said, cosmic strings arise naturally as stable topological defects in many field theory models and, depending on the model considered, we can have global $U(1)$ strings \cite{Vilenkin:1982ks}, the simplest case, Abelian-Higgs strings \cite{Nielsen:1973cs}, where the $U(1)$ symmetry is local, non-Abelian strings \cite{Bucher:1990gs,Alford:1992yx}, if the $K$ subgroup is non-Abelian, superconducting strings \cite{Witten:1984eb} and Nambu-Goto strings, in which the Langrangian is simplified to ignore string-core field excitation contributions to the string dynamics. In this work we will focus on an approximation of local $U(1)$ cosmic strings, also known as Abelian Higgs strings, following the prescription by Vilenkin\cite{Vilenkin:1981zs}.

Furthermore, cosmic string configurations can be obtained through the general relativistic field equations. In this case, we can assume a cylindrically symmetric metric of the type \cite{Christensen:1999wb}:
\begin{equation}
ds^2=-N^2(r)dt^2+dr^2+L^2(r)d\theta ^2+K^2(r)dz^2 \,,
\end{equation}
where $N$, $L$ and $K$ are arbitrary functions of the radial coordinate $r$. If one considers $N=K$, and the regularity conditions $L(0)=N'(0)=0$ and $N(0)=L'(0)=1$, a complete classification of the string-like solutions of the gravitating Abelian-Higgs model can be achieved \cite{Christensen:1999wb}.

The metric for a straight cosmic string oriented along the $z$-axis can be written in a quite general form as \cite{vandeMeent:2012gb}
\begin{equation}
\label{generalmetric}
ds^2=-dt^2+dr^2+ \left[1-\frac{\mu(r)}{2\pi}\right]^2r^2 d\theta ^2+dz^2 \,,
\end{equation}
where the Riemann curvature is zero except for the $tz$-hyperplane.
An interesting solution that arises from the linear approximation of General Relativity (GR) is the case where in Eq. \eqref{generalmetric} the non-trivial element of the metric takes the form $(1 - 8\pi G \mu)r^2$ \cite{Vilenkin:1981zs}, with $\mu$ being the linear density of the string. In this case the geometry of a massive string can be described as a conical singularity with a deficit angle proportional to the linear energy density, $2\pi-8\pi G\mu$.
In fact, an interesting property of this solution is that, in terms of the modified azimuthal coordinate $\theta ' =(1-4 G\mu)\theta$, ranging from 0 to $2\pi-8\pi G \mu$, the geometry is Minkowski, which implies that the gravitational acceleration of massive objects towards the string is zero \cite{Vilenkin:1981zs}. Generalisations of Eq. \eqref{lagrangiangoldstone} including self-coupling scalar fields can also be found in the literature \cite{Barriola:1989hx}.

Despite being  a fundamental feature on several grand unified scenarios \cite{Jeannerot:2003qv} and ubiquitous in many areas of condensed-matter physics, namely, in metal crystallization \cite{Mermin:1979zz}, liquid crystals \cite{1975JMoSt..29..190J,Chuang:1991zz}, superfluid helium-3 \cite{Salomaa:1987zz} and helium-4 \cite{1994Natur.368..315H}, and superconductivity \cite{Abrikosov:1956sx}, topological defects  still elude detection on cosmological levels. However, they remain an active and important field of study in Cosmology not only for the possibility of constraining the different GUT scenarios, but also for their possible impact on the cosmic microwave background (CMB) anisotropies  \cite{Ade:2013xla}, small structure formation \cite{Wu:1998mr}, reionization history \cite{Olum:2006at}, gravitational lensing observations \cite{Thomas:2009bm}, super-massive black-hole formation \cite{Lake:2015ppa} and gravitational wave (GW) spectra \cite{Binetruy:2012ze}.

String-type solutions have also been studied in the framework of modified theories of gravity, such as in Brans-Dicke theory \cite{Sen:1997pu}, general scalar-tensor theories \cite{Gundlach:1990nm,Barros:1994km,Guimaraes:1996ti,Boisseau:1997st,Dahia:1998vj,Sen:1997pu,Arazi:2000tn,Gregory:1997wk,Sen:1997hn,Sen:1998jj,Delice:2006gs}, $f(R,L_m)$ gravity \cite{Harko:2010mv,Harko:2014axa}, and hybrid metric-Palatini gravity (HMPG) \cite{Harko:2020oxq}.
In this work, we generalize the previous analysis and consider cosmic string-like objects in the generalized HMPG theory \cite{Tamanini:2013ltp}, which is described by an action depending on a general function of both the metric Ricci scalar $R$ and the Palatini Ricci scalar $\mathcal R$ written in terms of an independent connection $\hat\Gamma^c_{ab}$, thus adding an extra degree of freedom to the non-generalized hybrid metric-Palatini gravity on which the action is constructed from the Einstein-Hilbert action via the addition of an arbitrary function of $\mathcal R$ only.

The main motivation to investigate hybrid metric-Palatini theories resides on the fact that these theories are able to overcome flaws of both the metric and the Palatini approaches to $f\left(R\right)$ gravity. For example, in both the metric and the Palatini formalisms of $f\left(R\right)$ gravity, one is able to model the late-time cosmic acceleration period without invoking dark energy sources \cite{Sotiriou:2008rp}, but both approaches present unavoidable disadvantages: the metric $f\left(R\right)$ was shown to be inconsistent with solar-system constraints unless chameleon mechanisms are considered \cite{Khoury:2003aq,Khoury:2003rn}, whereas the Palatini $f\left(R\right)$ gravity induces microscopic instabilities, surface singularities in polytropic star models, and is unable to describe the evolution of cosmological perturbations \cite{Olmo:2011uz, Gomez:2020rnq}. The HMPG successfully unifies the late-time cosmic acceleration period with the weak-field solar system dynamics without the need for chameleon mechanisms \cite{Harko:2011nh}, thus being a viable and relevant modification to GR.

The applicability range of HMPG is immense and the theory was analyzed in a wide variety of areas of gravitational physics. In particular, the effects of a light long-range scalar field predicted by the theory were analyzed both in cosmological \cite{Capozziello:2012ny,Carloni:2015bua} and galactic \cite{Capozziello:2012qt,Capozziello:2013yha} dynamics, always with the advantage of being consistent with solar system constraints \cite{Capozziello:2013uya,Capozziello:2013wq,Dyadina:2019dsu}. Various scenarios were explored in a cosmological context \cite{Boehmer:2013oxa,Santos:2016tds,Kausar:2019iwu,Sa:2020qfd,Sa:2020fvn,Paliathanasis:2020fyp,Rosa:2017jld,Rosa:2019ejh,Rosa:2021ish}, observational constraints were discussed \cite{Lima:2014aza,Lima:2015nma,Leanizbarrutia:2017xyd,Avdeev:2020jqo}, and stability issues were addressed \cite{Koivisto:2013kwa,Capozziello:2013gza,Chen:2020evr}. Solutions describing compact objects were also explored, such as black-holes \cite{Danila:2018xya,Bronnikov:2019ugl,Bronnikov:2020vgg}, for which the Kerr black-hole was shown to be stable against scalar perturbations \cite{Rosa:2020uoi}, stars \cite{Danila:2016lqx}, and wormholes geometries \cite{Capozziello:2012hr,KordZangeneh:2020ixt}, some of which satisfying the null energy condition throughout the whole spacetime \cite{Rosa:2018jwp,Rosa:2021yym}. The weak-field regime \cite{Rosa:2021lhc} of the theory was also investigated with implications for gravitational wave physics \cite{Kausar:2018ipo,Bombacigno:2019did} and higher-dimensional braneworld scenarios \cite{Fu:2016szo,Rosa:2020uli} were also explored. We refer the reader to Refs. \cite{Harko:2018ayt,Capozziello:2015lza,Harko:2020ibn} for recent reviews on the topic.

\subsection{Outline of the paper}

This work is organised in the following manner: In Sec. \ref{secII}, we explicitly present the scalar-tensor representation of generalized HMPG, by writing the action and the gravitational field equations in a straight infinite cosmic string background, taking into account Vilenkin’s prescription and noting that local gauge strings preserve boost invariance. In Sec. \ref{secIII}, we consider exact cosmic string solutions, by considering two simple choices for the potential, the null potential and the constant potential. By considering more complicated functional forms of the potential one must resort to numerical methods in order to construct cosmic string models. In Sec. \ref{secIV}, we consider potentials with different structures: independent of $\psi$ and of second and forth order in $\xi$, independent of $\xi$ and second order on $\phi$, second order on both scalar fields and finally with an additive and a multiplicative structure, respectively, for the potential and, by varying the potential parameters and initial conditions, show that the numerical solutions strongly depend on the boundary values of the geometrical and physical parameters used to integrate the system of the field equations. Finally, in Sec. \ref{sec:conclusion}, we conclude and discuss our results.

\section{Cosmic strings in the scalar-tensor representation of generalized HMPG}\label{secII}

In the present Section, we introduce the action and the field equations of the generalized HMPG theory, and write down the system of equations describing cosmic strings in static cylindrical symmetry.

\subsection{Action and field equations}

The generalized HMPG theory is described by an action $S$ of the form
\be\label{genac}
S=\frac{1}{2\kappa^2}\int_\Omega\sqrt{-g}f\left(R,\cal{R}\right)d^4x+
\int_\Omega\sqrt{-g}\;{\cal L}_m d^4x,
\ee
where $\kappa^2 \equiv 8\pi G/c^4$, $G$ is the gravitational constant and $c$ the speed of light, $\Omega$ is the spacetime manifold described by a system of coordinates $x^a$, $g$ is the determinant of the spacetime metric $g_{ab}$, where Latin indices run from 0 to 3, $R=g^{ab}R_{ab}$ is the Ricci scalar of the metric $g_{ab}$ and where $R_{ab}$ is the Ricci tensor, $\mathcal{R}\equiv\mathcal{R}^{ab}g_{ab}$ is the Palatini Ricci scalar, where the Palatini Ricci tensor $\mathcal R_{ab}$ is defined in terms of an independent connection $\hat\Gamma^c_{ab}$ in the usual form as $\mathcal{R}_{ab}=\partial_c\hat\Gamma^c_{ab}-\partial_b\hat\Gamma^c_{ac}+\hat\Gamma^c_{cd}\hat\Gamma^d_{ab}-\hat\Gamma^c_{ad}\hat\Gamma^d_{cb}$, where $\partial_a$ denotes partial derivatives with respect to the coordinates $x^a$,
$f\left(R,\cal{R}\right)$ is a well-behaved function of $R$ and $\cal{R}$, and ${\cal L}_m$ is the matter Lagrangian density considered to be minimally coupled to the metric $g_{ab}$. Equation \eqref{genac} depends on two independent variables, namely the metric $g_{ab}$ and the independent connection $\hat\Gamma^c_{ab}$, and thus two equations of motion can be obtained.

Taking a variation of Eq. \eqref{genac} with respect to the metric $g_{ab}$ leads to the modified field equations
\beq
\frac{\partial f}{\partial R}R_{ab}+\frac{\partial f}{\partial \mathcal{R}}\mathcal{R}_{ab}-\frac{1}{2}g_{ab}f\left(R,\cal{R}\right)
-\left(\nabla_a\nabla_b-g_{ab}\Box\right)\frac{\partial f}{\partial R}=\kappa^2 T_{ab},\label{field1}
\eeq
where $\nabla_a$ denotes covariant derivatives and $\Box=\nabla^a\nabla_a$ the d'Alembert operator, both with respect to $g_{ab}$, and $T_{ab}$ is the energy-momentum tensor defined as usual:
\begin{equation}
T_{ab}=-\frac{2}{\sqrt{-g}}\frac{\delta(\sqrt{-g}\,{\cal L}_m)}{\delta(g^{ab})} ~.
 \label{defSET}
\end{equation}
On the other hand, taking a variation of Eq. \eqref{genac} with respect to the independent connection $\hat\Gamma^c_{ab}$ yields
\be
\hat\nabla_c\left(\sqrt{-g}\frac{\partial f}{\partial \cal{R}}g^{ab}\right)=0 \,,
\label{eqvar1}
\ee
where $\hat\nabla_a$ is the covariant derivative written in terms of the independent connection $\hat\Gamma^c_{ab}$. Since $\sqrt{-g}$ is a scalar density of weight 1, then $\hat\nabla_c \sqrt{-g}=0$ and Eq. \eqref{eqvar1} can be rewritten in the form $\hat\nabla_c\left(\frac{\partial f}{\partial \cal{R}}g^{ab}\right)=0$. This result implies the existence of a new metric, $h_{ab}$, conformally related to the metric $g_{ab}$ via
\be
h_{ab}=g_{ab} \frac{\partial f}{\partial \cal{R}} \,,
\label{hab}
\ee
in such a way that the independent connection is the Levi-Civita connection of the metric $h_{ab}$, i.e., $\hat\Gamma^c_{ab}$ can be written as
\be
\hat\Gamma^a_{bc}=\frac{1}{2}h^{ad}\left(\partial_b h_{dc}+\partial_c h_{bd}-\partial_d h_{bc}\right)\,.
\ee

\subsection{Scalar-tensor representation of generalized HMPG with matter}\label{str}

It is sometimes useful to represent the generalized HMPG theory in a dinamically equivalent scalar-tensor representation in which the two extra scalar degrees of freedom of the theory are explicitly carried by two scalar fields. To obtain this representation, we introduce two auxiliary fields $\alpha$ and $\beta$ into Eq. \eqref{genac} and rewrite it in the form
\beq
S=\frac{1}{2\kappa^2}\int_\Omega \sqrt{-g}\left[f\left(\alpha,\beta\right)+\frac{\partial f}{\partial \alpha}\left(R-\alpha\right)
+\frac{\partial f}{\partial\beta}\left(\cal{R}-\beta\right)\right]d^4x+\int_\Omega\sqrt{-g}\;{\cal L}_m d^4x.
\label{gensca}
\eeq
At this point one verifies that if $\alpha=R$ and $\beta=\mathcal R$ one recovers Eq. \eqref{genac}. This equivalence between the two representations is only guaranteed if the determinant of the Hessian matrix of the function $f\left(\alpha,\beta\right)$ is non-zero, i.e., if $f_{\alpha\alpha}f_{\beta\beta}-f_{\alpha\beta}^2\neq0$, where the subscripts $\alpha$ and $\beta$ denote partial derivatives with respect to these variables. Defining two scalar fields as $\varphi=\partial f(\alpha,\beta)/\partial\alpha$ and $\psi=-\partial f(\alpha,\beta)/\partial\beta$, where the negative sign in $\psi$ is imposed to avoid the presence of ghosts, Eq. \eqref{gensca} takes the form
\beq
S=\frac{1}{2\kappa^2}\int_\Omega \sqrt{-g}\left[\varphi R-\psi\mathcal{R}-V\left(\varphi,\psi\right)\right]d^4x
+\int_\Omega\sqrt{-g}\;{\cal L}_m d^4x,
  \label{action3}
\eeq
where the function $V\left(\varphi,\psi\right)$ plays the role of an interaction potential between the two scalar fields and it is defined as
\be\label{potential}
V\left(\varphi,\psi\right)=-f\left(\alpha,\beta\right)+\varphi\alpha-\psi\beta \,.
\ee
Recalling the conformal relation between $h_{ab}$ and $g_{ab}$ in Eq. \eqref{hab}, which can now be written in the form $h_{ab}=-\psi\, g_{ab}$ by taking into consideration the definition of $\psi$, one can derive a relationship between $R$ and $\mathcal R$ as
\be\label{confrt}
\mathcal{R}=R+\frac{3}{\psi^2}\partial^a \psi\partial_a \psi-
\frac{3}{\psi}\Box\psi\,,
\ee
which can be used to eliminate $\mathcal R$ from Eq. \eqref{action3} and gives the final form of the action
\beq\label{genacts2}
S=\frac{1}{2\kappa^2}\int_\Omega \sqrt{-g}\Big[\left(\varphi-\psi\right) R
-\frac{3}{2\psi}\partial^a\psi\partial_a\psi
 -V\left(\varphi,\psi\right)\big]d^4x+\int_\Omega\sqrt{-g}\;{\cal L}_m d^4x.
\eeq

Equation \eqref{genacts2} is now a function of three independent variables, namely, the metric $g_{ab}$ and the two scalar fields $\varphi$ and $\psi$. Taking a variation of Eq. \eqref{genacts2} with respect to the metric $g_{ab}$ yields the modified field equations in the scalar-tensor representation. Varying the action \eqref{genacts2} with respect to the metric $g_{ab}$
provides the following gravitational equation
\beq
\left(\varphi-\psi\right) G_{ab}=\kappa^2T_{ab}
+\nabla_a\nabla_b
\varphi-\nabla_a\nabla_b\psi
+\frac{3}{2\psi}\partial_a\psi\partial_b\psi
 -\left(\Box\varphi-\Box\psi+\frac{1}{2}V+\frac{3}{4\psi}
\partial^c\psi\partial_c\psi\right)g_{ab}\,.
\label{genein2}
\eeq
Note that Eq. \eqref{genein2} could be obtained directly from Eq. \eqref{field1} via the introduction of the definitions of $\varphi$, $\psi$ and $V\left(\varphi,\psi\right)$, which further emphasizes the equivalence between the two representations.

Finally, the equations of motion for the scalar fields $\varphi$ and $\psi$ can be obtained via a variation of Eq. \eqref{genacts2} with respect to these fields, respectively, which after algebraic manipulations can be written in the forms
\be\label{genkgi}
\Box\varphi+\frac{1}{3}\left(2V-\psi V_\psi-\varphi V_\varphi\right)
=\frac{\kappa^2T}{3}\,,
\ee
\be\label{genkg}
\Box\psi-\frac{1}{2\psi}\partial^a\psi\partial_a\psi-\frac{\psi}{3}
\left(V_\varphi+V_\psi\right)=0\,,
\ee
respectively.

Notice from Eq.\eqref{genacts2} that the coupling between the scalar fields and the Ricci scalar is the combination $\varphi-\psi$. Since $\varphi$ and $\psi$ are arbitrary functions, it is not guaranteed that this combination preserves the positivity of the coupling. We thus introduce a redefinition of the scalar field $\varphi$ as $\xi^2=\varphi-\psi$. With this redefinition, any solution obtained for which $\xi$ is a real function preserves the positivity of the coupling $\left(\varphi-\psi\right)R$. Equations \eqref{genein2} to \eqref{genkg} thus become
\begin{equation}\label{genein3}
\xi^2G_{ab}=\kappa^2 T_{ab}+\nabla_a\nabla_b\xi^2+\frac{3}{2\psi}\partial_a\psi\partial_b\psi-\left(\Box\xi^2+\frac{1}{2}\bar V+\frac{3}{4\psi}\partial^c\psi\partial_c\psi\right)g_{ab},
\end{equation}
\begin{equation}\label{genkgxi}
\Box\xi^2+\frac{1}{2\psi}\partial^a\psi\partial_a\psi+\frac{1}{6}\left(4\bar V-\xi \bar V_\xi\right)=\frac{\kappa^2T}{3},
\end{equation}
\begin{equation}\label{genkgpsi}
\Box\psi-\frac{1}{2\psi}\partial^a\psi\partial_a\psi-\frac{\psi}{3}\left(\frac{1}{2\xi}\bar V_\xi+\bar V_\psi\right)=0,
\end{equation}
where $\bar V\left(\xi,\psi\right)$ is the potential written in terms of the scalar fields $\xi$ and $\psi$ and the subscript $\xi$ denotes a partial derivative with respect to this scalar field. In the next section, we will use the equations of motion (\ref{genein3})--(\ref{genkgpsi}) to find cosmic string solutions. Finally, one can also obtain a relationship between the potential $\bar V$ and the function $f\left(R,\mathcal R\right)$ from Eq.\eqref{potential} as
\begin{equation}\label{Vin}
\bar V\left(\xi,\psi\right)=-f\left(R,\mathcal R\right)+\xi^2 R+\psi\left(R-\mathcal R\right),
\end{equation}
where we have used the fact that the scalar-tensor representation is only defined if $\alpha=R$ and $\beta=\mathcal R$. This equation becomes a PDE for $f\left(R,\mathcal R\right)$ by replacing $\psi=f_\mathcal R$ and $\xi^2=f_R+f_\mathcal R$.

Hence Eq.~(\ref{Vin}) becomes
\begin{equation}\label{Vin1}
V\left( \sqrt{\frac{\partial f\left( R,\mathcal{R}\right) }{\partial R}+\frac{\partial
f\left( R,\mathcal{R}\right) }{\partial \mathcal{R}}},\frac{\partial f\left( R,\mathcal{R}\right) }{%
\partial \mathcal{R}}\right) =-f\left( R,\mathcal{R}\right) +R \frac{\partial
f\left( R,\mathcal{R}\right) }{\partial R}+\mathcal R\frac{\partial f\left( R,\mathcal{R}\right) }{%
\partial \mathcal{R}}.
\end{equation}

\subsection{Metric of a cosmic string-like object}\label{metrcs}

Due to the existence of a magnetic flux inside the string \cite{Nielsen:1973cs}, cosmological strings can either be infinite or form closed loops, that will oscillate and decay emitting energy through GWs, originating a GW stochastic background \cite{Abbott:2017mem}.
The crucial parameter in the analysis of this type of string is, as seen before, the energy density of the string, usually represented as $G\mu$. This density, related to the energy scale of the symmetry breaking originating the network of strings, is constrained by different types of observations, ranging from the CMB spectra \cite{Battye:2010xz}, to gravitational lensing \cite{Mack:2007ae} and 21-cm observation \cite{Khatri:2008zw}, and more recently from the GW observations \cite{Abbott:2017mem}. In the future, we expect a more tight constrain coming from the GW spectra with LISA \cite{Caprini:2015zlo}.

In this work, we consider a straight infinite Abelian-Higgs cosmic string, using Vilenkin's approximation \cite{Vilenkin:1981zs}
\begin{equation}\label{string}
    T^t_t=T^z_z=-\sigma(r) \,,
\end{equation}
where $\sigma$ is the string tension.
We further assume cylindrical symmetry with a general metric of the form:
\begin{equation}\label{metric}
    ds^2=-e^{2(K-U)}dt^2+e^{2(K-U)}dr^2+e^{-2U}W^2d\theta^2+e^{2U}dz^2,
\end{equation}
where $t$, $r$, $\theta$ and $z$ denote the time, radial, angular and axial cylindrical coordinates, respectively, and $K$, $U$ and $W$ are functions of $r$ alone. This metric is invariant under a set of transformations of the type: $x^0 \rightarrow x^0 +c_1$, $x^3\rightarrow x^3 +c_2 $, $x^0 \rightarrow -x^0$ and $x^3 \rightarrow -x^3$, where $c_1$ and $c_2$ are constants, which renders this metric static and cylindrically symmetric.
Thus, taking into account Eqs. \eqref{string} and \eqref{metric}, we are assuming that the geometry of the string is symmetric with respect to rotations about the cylinder axis and also to translations along the axis direction.

\subsection{Field equations with boost invariance}

Since in this model the matter field couples minimally with curvature, it is possible to show that the energy conservation equation still holds, i.e.,
\begin{equation}\label{conservation}
    \nabla_a T^{a}{}_{b}=0
\end{equation}
which provides $K'\sigma=0$ and, apart from the trivial vacuum solution $\sigma=0$, this implies that $K'=0$, where the prime represents a differentiation w.r.t. $r$. Thus, we consider from now on that $e^K=1$.

Note that local gauge strings preserve boost invariance along $t$ and $z$ \cite{Vilenkin:1981zs}, so that this requires $U=0$. Hence the only surviving non-trivial metric tensor component is $g_{\theta \theta}=W^2(r)$.  From a geometric point of view $W(r)$ is the radius of the coordinate
circles $r={\rm constant}$, $z={\rm constant}$, parameterized by the angle $\theta$. Since in this geometry the perimeter of a circle equals $2\pi W$, in the following we will call the only remaining metric tensor component $W^2(r)$ a {\it circular radius}. On the other hand $W^2(r)$ also has the geometric meaning  of a length that may be counted from any zero point, with its value at $r=0$ not distinguished geometrically. Hence the metric of the cosmic string reduces to the form
\be\label{metrn}
ds^2=-dt^2+dr^2+W^2(r)d\theta^2+dz^2.
\ee

Applying this symmetry, the gravitational field equations simplify considerably. Equation \eqref{genein3} provides three independent field equations, which are
\begin{eqnarray}\label{fieldtt}
\xi^2\frac{W''}{W}+2\xi\xi'\frac{W'}{W}
+\frac{3\psi'^2}{4\psi}+2\left(\xi'^2+\xi\xi''\right)+\frac{\bar V}{2}=-\kappa^2\sigma,
\end{eqnarray}
\begin{equation}\label{field11}
2\xi\xi'\frac{W'}{W}-\frac{3\psi'^2}{4\psi}+\frac{\bar V}{2}=0,
\end{equation}
\begin{equation}\label{field22}
2\left(\xi'^2+\xi\xi''\right)+\frac{3\psi'^2}{4\psi}+\frac{\bar V}{2}=\frac{d^2}{dr^2}\xi^2+\frac{3\psi'^2}{4\psi}+\frac{\bar V}{2}=0 \,,
\end{equation}
whereas the scalar field equations for $\xi$ and $\psi$, provided by Eqs. \eqref{genkgxi} and \eqref{genkgpsi}, give
\begin{equation}\label{eomphi}
2\left(\xi'^2+\xi\xi''\right)+2\xi\xi'\frac{W'}{W}+\frac{\psi'^2}{2\psi}+\frac{1}{6}\left(\bar V-\xi \bar V_\xi\right)=-\frac{2\kappa^2}{3}\sigma,
\end{equation}
\begin{equation}\label{eompsi}
\psi''+\frac{W'}{W}\psi'-\frac{\psi'^2}{2\psi}-\frac{\psi}{3}\left(\bar V_\psi+\frac{1}{2\xi}\bar V_\xi\right)=0 \,.
\end{equation}

The system of Eqs. \eqref{fieldtt}-\eqref{eompsi} is a system of five equations from which only four are linearly independent. This statement can be proven by taking a radial derivative of Eq. \eqref{field11}, using Eq. \eqref{fieldtt} to cancel the factors $W''$, using Eq. \eqref{eomphi} to cancel $\sigma$, and finally using Eqs. \eqref{field22} and \eqref{eompsi} to cancel the second-order derivatives of the scalar fields $\xi''$ and $\psi''$, respectively. As a result, one recovers Eq. \eqref{field11}, thus proving that the system of equations is linearly dependent. Thus, one only needs to consider four of these equations to completely determine the solution in the sections that follow. Given its complexity, we chose to discard Eq. \eqref{fieldtt} from the analysis.

Furthermore, an equation for the potential $\bar V$ can be obtained by summing the field equations in Eqs. \eqref{field11} and \eqref{field22}, yielding
\begin{equation}\label{eq28}
\bar V=-2\left(\xi'^2+\xi\xi''\right)-2\xi\xi'\frac{W'}{W}.
\end{equation}
This equation is particularly useful to obtain an equation for $W'$ in terms of the scalar fields $\xi$ and $\psi$ and their derivatives after setting an explicit form of the potential $\bar V$, which we explore below in Sec. \ref{secIII}.

The system of basic equations describing the structure of a cosmic string can thus be reformulated in the form of a first-order dynamical system. By defining $\alpha =\xi ^2$, and  by introducing two extra dynamical variables as $u=\alpha '$ and $v=\psi'$, the dynamical system takes the form
\begin{equation}\label{eq1}
\frac{d\alpha }{dr}=u, \qquad \frac{d\psi }{dr}=v,
\end{equation}%
\begin{equation}\label{eq2}
\frac{dW}{dr}=\frac{1}{u}\left( \frac{3v^{2}}{4\psi }-\frac{\bar{V}}{2}\right) W,
\end{equation}%
\begin{equation}\label{eq3}
\frac{du}{dr}=-\frac{3v^2}{4\psi}-\frac{\bar{V}}{2},
\end{equation}%
\begin{equation}\label{eq4}
\frac{dv}{dr}=-\frac{v}{u}\left(\frac{3v^2}{4\psi}-\frac{\bar{V}}{2}\right)+\frac{v^2}{2\psi}+\frac{\psi }{3}\left( \bar{V}_{\psi }+\frac{1}{2\sqrt{\alpha}}\bar{V}_{\sqrt{\alpha} }\right) ,
\end{equation}
where Eq. \eqref{eq1} is the explicit definition of $u$ and $v$, and Eqs. \eqref{eq2}--\eqref{eq4} are reformulations of Eqs. \eqref{field11}, \eqref{eomphi}, and \eqref{eompsi}, respectively. Once the functional form of the potential $\bar{V}(\xi,\psi)$ is specified, the system of Eqs. (\ref{eq1})-(\ref{eq4}) represents a system of ordinary, strongly nonlinear, differential equations for the variables $\left(\alpha =\xi ^2,\psi, W, u,v\right)$. To solve this system, one has to impose a set of boundary conditions at some radius $r=r_0$, i.e., $\alpha \left(r_0\right)=\alpha _0$, $\psi \left(r_0\right)=\psi _0$, $W \left(r_0\right)=W _0$, $u \left(r_0\right)=u _0$, and $v \left(r_0\right)=v_0$, respectively, which specify the boundary values of the variables on, or nearby the string axis. Moreover, we will also impose the condition $u\left(r_0\right)\neq v\left(r_0\right)$. Once the system is solved, the string tension can be obtained from Eq.~(\ref{eomphi}), and it is given by
\be
\frac{2}{3}\kappa ^2 \sigma=\bar{V}-\frac{v^2}{2\psi}-\frac{1}{6}\left(\bar{V}-\sqrt{\alpha}\bar{V}_{\sqrt{\alpha}}\right).
\ee

An important physical characteristic of the string-like objects is their mass per unit length $m_s$, defined as
\beq
m_s\left(R_s\right)=\int_0^{2\pi}{d\theta}\int_0^{R_s}{\sigma (r)W(r)dr}
	= 2\pi \int_0^{R_s}{\sigma (r)W(r)dr},
\eeq
where $R_s$ is the radius of the string, defined as the distance from the center where the string tension vanishes, $\sigma \left(R_s\right)=0$, and $\sigma (r)\equiv 0, \forall r\geq R_s$. Note that, in general, the solutions obtained for $\sigma$ do not satisfy the property $\sigma (r)\equiv 0, \forall r\geq R_s$, and this condition must be imposed manually by performing a matching between the string spacetime and an exterior cosmological spacetime. This matching must be performed via the use of the junction conditions of the theory, previously used in \cite{Rosa:2018jwp}. However, we do not pursue this analysis here as it is out of the scope of this paper. Using Eqs.~(\ref{field22}) and Eq.~(\ref{fieldtt}) the mass per unit length of the string can be expressed as
\be
\kappa ^2 m_s\left(R_s\right)=3\pi\int_0^{R_s}{\left[\bar{V}-\frac{v^2}{2\psi}-\frac{1}{6}\left(\bar{V}-\sqrt{\alpha}\bar{V}_{\sqrt{\alpha}}\right)\right]Wdr}.
\ee

\section{Exact cosmic string solutions}\label{secIII}

The solutions of the gravitational field equations for a cosmic string-like configuration in generalized HMPG essentially depend on the functional form of the potential $\bar{V}=\bar{V}(\xi,\psi)$. Once the potential and the boundary conditions are given,  solutions of the gravitational field equations can be obtained that uniquely fix the metric tensor component $W$, as well as the two scalar fields $\xi$ and $\psi$. In the following, we will consider first several exact solutions of the field equations, corresponding to two simple choices of the potential $V$.

\subsection{$\bar{V}(\xi,\psi)=0$}

We begin our investigations of the string-like structures in generalized HMPG by considering the simple case in which the potential $\bar{V}$ vanishes, $\bar{V}=0$. In this case Eq.~(\ref{Vin1}) takes the form
\be
-f\left( R,\mathcal{R}\right) +R \frac{\partial
f\left( R,\mathcal{R}\right) }{\partial R}+\mathcal R\frac{\partial f\left( R,\mathcal{R}\right) }{%
\partial \mathcal{R}}=0,
\ee
and it has the general solution
\be
f\left( R,\mathcal{R}\right)=R g\left(\frac{\mathcal R}{R}\right)+\mathcal R h\left(\frac{R}{\mathcal R}\right),
\ee
where $g$ and $h$ are arbitrary functions.

From Eq.~(\ref{eq28}) we obtain
\be
\frac{W'}{W}=-\frac{\left(\xi ^2\right)''}{\left(\xi ^2\right)'},
\label{w1}
\ee
giving immediately
\be
W=\frac{W_0}{\left(\xi ^2\right)'}=\frac{W_0}{2\xi \xi '},
\ee
where $W_0$ is an integration constant. On the other hand, from Eq.~(\ref{eompsi}) we obtain
\be\label{w2}
\frac{\psi''}{\psi'}+\frac{W'}{W}-\frac{1}{2}\frac{\psi '}{\psi}=0,
\ee
and
\be\label{w2}
 W=\frac{C\sqrt{\psi}}{\psi '},
\ee
respectively, where $C$ is an arbitrary integration constant. Eqs~(\ref{w1}) and (\ref{w2}) can be combined to obtain
\be\label{eq39}
\frac{d}{dr}\xi ^2=\frac{w_0\psi '}{\sqrt{\psi}}=2w_0\frac{d}{dr}\sqrt{\psi},
\ee
where we have defined $w_0=W_0/C$. Taking the radial derivative of Eq. \eqref{eq39} and inserting the result into Eq. \eqref{field22} yields an ODE for $\psi$ in the form
\be\label{eqpsiV0}
\psi''-\frac{1}{2}\frac{\psi '^2}{\psi}+\frac{3}{4w_0}\frac{\psi '^2}{\sqrt{\psi}}=0.
\ee
Dividing through by $\psi'$ allows one to integrate Eq. \eqref{eqpsiV0} directly twice and obtain the solution for $\psi\left(r\right)$ in the form
\be\label{eq45}
\psi (r)=\frac{4}{9} w_0^2 \ln
   ^2\left[\frac{3 c_1 \left(r+c_2\right)}{4 w_0}\right],
\ee
where $c_1$ and $c_2$ are arbitrary integration constants. From Eq.~(\ref{eq39}) we obtain
\be
\xi ^2=\xi _0^2+\frac{4}{3}w_0^2\ln \frac{3c_1\left(r+c_2\right)}{4w_0}.
\ee
Thus we find for the metric tensor component $W$ the expression
\be\label{w1}
W(r)=\frac{3 C^2 \left(r+c_2\right)}{4 W_0}.
\ee
Finally, the string tension can be computed directly from Eq. \eqref{eomphi} and it is given by
\be
-\kappa ^2\sigma =\frac{4 W_0^2}{3C^2 \left(c_2+r\right)^2}.
\ee

Note that the above solutions have been obtained under the assumption $\ln \left[3 c_1 \left(c_2+r\right)/4
   w_0\right] >0$. By assuming that the integration constant $c_2 \neq 0$, it follows that $W$ takes on the string axis the boundary value $W(0)=3c_1c_2/4w_0$. However, for the choice $c_2=0$, $W$ vanishes on the string axis. But, on the other hand, this choice would imply a divergent string tension and scalar fields at $r=0$. The circular radius $W(r)$ monotonically increases with the radial distance from the center, and for $r\rightarrow \infty$, $\lim_{r\rightarrow \infty}W^2(r)=\infty$.  In the same limit the string tension tends to zero, indicating a vanishing string tension at infinity.

An interesting property of the present zero potential solution for cosmic strings in generalized HMPG theory is that the string tension is negative. In \cite{Visser:1989kh} wormhole configurations representing an alternative of the cosmic string solutions of standard GR have been obtained under the assumption of a negative string tension and mass. The properties of such configurations have been further investigated in \cite{Cramer:1994qj}, where it was pointed out that a wormhole mouth embedded in high background matter  density, and which accretes mass, can give the other mouth a net negative mass. The lensing of such
gravitationally negative anomalous objects will have observable lensing properties. However, for the present vanishing potential generalized HMPG theory cosmic string solution, the mass of the string is given by
\be
m_s\left(R_s\right)=\frac{2 \pi W_0}{\kappa ^2} \ln \frac{c_2}{c_2+R_s}.
\ee
 For $R_s\rightarrow \infty$, the mass of the cosmic string diverges logarithmically to minus infinity.

 For $\sigma =0$, that is, in the vacuum, the field equations \eqref{fieldtt}-\eqref{eompsi} do admit the simple solution
 \be
 \psi=\psi _0={\rm constant},W={\cal W}_0={\rm constant}, \xi ^2=c_3r+c_4,
 \ee
where $c_1$ and $c_2$ are integration constants. The corresponding vacuum metric is given by
\be
ds^2=-dt^2+dr^2+{\cal W}_0^2d\theta^2+dz^2.
\ee

A matching of this vacuum metric with the string metric tensor component $W(r)$ at a finite radius $R_s$ determines the string radius as
 \be
 R_s=\frac{4{\cal W}_0W_0}{3C^2}-c_2.
 \ee
 The string tension takes then the surface value
 \be
 \kappa ^2 \sigma =\frac{3}{4}\frac{C^2}{{\cal W}_0^2}.
 \ee

 Hence there is a sudden transition from a finite (negative) value of the string tension to its zero vacuum value.

\subsection{$V(\xi,\psi)=\Lambda={\rm constant}$}

Next we consider the case when the potential $V$ is a constant, so that $V=\Lambda={\rm constant}$.  In this case Eq.~(\ref{Vin1}) takes the form
\be
-f\left( R,\mathcal{R}\right) +R \frac{\partial
f\left( R,\mathcal{R}\right) }{\partial R}+\mathcal R\frac{\partial f\left( R,\mathcal{R}\right) }{%
\partial \mathcal{R}}=\Lambda,
\ee
and it has the general solution
\be
f\left( R,\mathcal{R}\right)=R g\left(\frac{\mathcal R}{R}\right)+\mathcal R h\left(\frac{R}{\mathcal R}\right)-\Lambda,
\ee
where $g$ and $h$ are arbitrary functions.

For a constant potential  Eq.~(\ref{eompsi}) simplifies to
\be\label{eq51}
\frac{W'}{W}=-\frac{\psi''}{\psi '}+\frac{1}{2}\frac{\psi '}{\psi},
\ee
which allows us to write Eq.~(\ref{field11}) in the form
\beq
\hspace{-0.5cm}\frac{d}{dr}\xi ^2&=&\frac{3\psi '^2/4\psi-\Lambda/2}{W'/W}=\frac{3\psi '^2/4\psi-\Lambda/2}{-\psi ''/\psi'+(1/2)\psi'/\psi}.\label{eqVL}
\eeq
To facilitate the analysis, we introduce now a new function $h=\psi '^2/\psi$. The radial derivative of this function can be written in terms of $\psi$ and its derivatives as
  \be
  h'=2h\left(\frac{\psi ''}{\psi '}-\frac{1}{2}\frac{\psi '}{\psi}\right).
  \ee

This definition allows us to rewrite Eq. \eqref{eqVL} in terms of $h$ as
  \be\label{eq55}
 \frac{d}{dr}\xi ^2=-\frac{(3/4)h-\Lambda/2}{h'/2h},
  \ee
which can then be differentiated with respect to $r$ and inserted into Eq.~(\ref{field22}) to cancel the dependency in $d^2\xi ^2/dr^2$ and $\psi''$. As a result, we obtain an equation depending solely in $h$ of the form
\be\label{eq58}
\frac{(2 \Lambda -3 h) \left(3 h'^2-2 h h''\right)}{4 h'^2}=0,
\ee
This equation is undefined for $h'=0$, as the denominator vanishes in this case. Thus, in the following we ignore the solution corresponding to $h=2\Lambda/3={\rm constant}$, giving $h'=0$, and $W={\rm constant}$. The general solution of Eq.~(\ref{eq58}) is given by
\be\label{solh}
h(r)=\frac{c_2}{\left(r+2
   c_1\right)^2},
\ee
where $c_1$ and $c_2$ are arbitrary integration constants. Recalling that $h'=\psi'^2/\psi$, Eq.~\eqref{solh} becomes a separable ODE for $\psi$ which can be directly integrated and provides the general solution
\be\label{solpsiVL}
\psi (r)=\left[c_3\pm \frac{\sqrt{c_2}}{2}\ln \left(r+2c_1\right)\right]^2,
\ee
where $c_3$ is an arbitrary integration constant. For $c_3=0$, we recover the expression (\ref{eq45}) of $\psi (r)$, corresponding to the case $V=0$.  Inserting Eq.~\eqref{solpsiVL} into Eq.~(\ref{eq55}) and integrating gives for $\xi ^2$ the expression
\beq
\xi^2 (r)=\xi _0^2
	+\frac{1}{4} \left[3 c_2 \ln \left(r+2 c_1\right)-\Lambda  r \left(r+4 c_1\right)\right],
\eeq
where $\xi _0^2$ is an integration constant. Inserting Eq. \eqref{solpsiVL} into Eq. \eqref{eq51} we obtain the solution for $W$
\be
W(r)=W_0\left(r+2c_1\right),
\ee
where $W_0$ is a constant of integration. Hence the cosmic string metric tensor component $W(r)$ is the same in both $V=0$ and $V=\Lambda$ cases. Finally, the string tension can be computed via Eq. \eqref{eomphi}, leading to
\be
\kappa ^2\sigma =\frac{\Lambda }{2}-\frac{3 c_2}{4 \left(2 c_1+r\right)^2}.
\ee

On the string axis $r=0$ we obtain for the string tension the value
\be
\kappa ^2\sigma _0=\kappa ^2 \sigma (0)=\frac{\Lambda}{2}-\frac{3c_2}{16c_1^2}.
\ee
The condition of the positivity of the string tension imposes the condition $c_2/c_1^2<8\Lambda /3$ on the integration constants.

For $\Lambda =0$ we reobtain the expression corresponding to the case $V=0$. However, by an appropriate choice of the integration constants, and by assuming $\Lambda >0$, the string tension can be made positive in this model for all $r>0$. Moreover, $\lim_{r\rightarrow \infty}\sigma (r)=\Lambda/2\kappa ^2$, and hence at infinity the string tension becomes equal to the cosmological constant. However, in this case one can obtain a finite radius string configuration, with the radius $R_s$ determined by the condition $\sigma \left(R_s\right)=0$, and given by
\be
R_s=\sqrt{\frac{3}{2}\frac{c_2}{\Lambda}}-2c_1.
\ee
For a positive string tension at the origin $r=0$, the string radius is also positive.

As for the mass of the string we obtain
\beq
m_s\left(R_s\right)=\frac{2\pi W_{0}}{\kappa ^{2}}\Bigg\{ \frac{1}{2}\Lambda \left[
R_{s}+c_{1}(\Lambda W_{0}+2)-\frac{3c_{2}W_{0}}{8c_{1}}\right]
	- \frac{%
6c_{1}c_{2}}{8c_{1}\left[ R_{s}+c_{1}(\Lambda W_{0}+2)\right] -3c_{2}W_{0}}%
\Bigg\}.
\eeq
By an appropriate choice of the integration constants, giving the boundary conditions of the fields $\varphi$ and $\psi$ for $r=0$, one can always satisfy the condition $m_s\left(R_s\right)>0, \forall R_s$. In the limit $R_s\rightarrow \infty$, we obtain $m_s\left(R_s\right)\approx \left(\pi W_0\Lambda/\kappa  ^2\right)R_s$, that is, for large distances the mass of the string linearly increases with its radius.

\section{Numerical cosmic string solutions}\label{secIV}

Except for the simple cases considered in the previous Section, for potentials having more complicated functional forms one must resort to numerical methods in order to construct cosmic string models in the generalized HMPG theory. In the present Section, we obtain a number of such numerical solutions describing cosmic string configurations, by fixing first the form of the potential. We will adopt several forms for $V(\xi, \psi)$, by assuming that it has an additive, and a multiplicative structure, respectively. Moreover, we will also consider the cases where $V$ depends on only one variable, $\xi$ and $psi$, respectively, so that $V=V(\xi)$ and $V=V(\psi)$, respectively. The numerical solutions strongly depend on the boundary values of the geometrical and physical parameters used to integrate the  system of equations (\ref{eq1})--(\ref{eq4}), which, at least in the present approach, can be chosen arbitrarily. Moreover, the potential also depends on some numerical parameters. Hence, in the following, we will study two types of different effects on the numerical cosmic string configurations, related to the variation of the potential parameters, and of the boundary conditions, respectively.

\subsection{$\bar{V}(\xi,\psi)=\bar{V}(\xi)$}

We begin our numerical investigations by first considering the case in which the potential $\bar{V}\left(\xi, \psi\right)$ is independent on the variable $\psi$, $\bar{V}=\bar{V}(\xi)$. We will study only potentials $V(\xi)$ that have a simple polynomial dependence on $\xi$, $\bar{V}=\bar{V}_0\xi ^n$, and for the sake of concreteness we will restrict our analysis to the cases $n=2$ and $n=4$, respectively. These forms of the potential are chosen because the associated forms of the function $f\left(R,\mathcal R\right)$ have been proven to provide interesting cosmological behaviors in other works \cite{Rosa:2017jld}, e.g., for $n=2$ one can obtain the de-Sitter solution as well as cosmological bounces, and for $n=4$ one can reproduce the matter-dominated era.

Given the dependence of the scalar field $\xi$ in both the derivatives of $f\left(R,\mathcal R\right)$ with respect to $R$ and $\mathcal R$, the symmetric structure of Eq.~(\ref{Vin1}) when the potential $V$ depends solely on $\xi$ allows for the analytic derivation of families of solutions for the function $f\left(R,\mathcal R\right)$. In the following sections, when one considers potentials that depend also on $\psi$, this symmetry is broken and one can only obtain particular solutions of Eq.~(\ref{Vin1}).

\subsubsection{$\bar{V}(\xi,\psi)=\bar{V}_0\xi^2$}

We assume now that $\bar{V}$ is a simple quadratic power law function, so that $\bar{V}=\bar{V}_0\xi^2$. In this case, Eq.~(\ref{Vin1}) becomes
\begin{equation}
-f\left( R,\mathcal{R}\right) +R \frac{\partial
f\left( R,\mathcal{R}\right) }{\partial R}+\mathcal R\frac{\partial f\left( R,\mathcal{R}\right) }{%
\partial \mathcal{R}}=\bar V_0\left(\frac{\partial
f\left( R,\mathcal{R}\right) }{\partial R}+\frac{\partial f\left( R,\mathcal{R}\right) }{%
\partial \mathcal{R}}\right),
\end{equation}
which can be solved with respect to $f\left(R,\mathcal R\right)$ to obtain
\begin{equation}
f\left(R,\mathcal R\right)=\left(R-\bar V_0\right)g\left(\frac{\mathcal R-\bar V_0}{R-\bar V_0}\right)+\left(\mathcal R-\bar V_0\right)h\left(\frac{R-\bar V_0}{\mathcal R-\bar V_0}\right),
\end{equation}
where $g$ and $h$ are arbitrary functions.

 The system of field equations to be solved takes the form
\begin{equation}\label{eq1a}
\frac{d\alpha }{dr}=u, \qquad \frac{d\psi }{dr}=v,
\end{equation}%
\begin{equation}\label{eq2a}
\frac{dW}{dr}=\frac{1}{2u}\left( \frac{3v^{2}}{2\psi }-\bar{V}_0\alpha\right) W,
\end{equation}%
\begin{equation}\label{eq3a}
\frac{du}{dr}=-\frac{3v^2}{4\psi}-\frac{\bar{V}_0\alpha}{2},
\end{equation}%
\begin{equation}\label{eq4a}
\frac{dv}{dr}=-\frac{v}{u}\left(\frac{3v^2}{4\psi}-\frac{\bar{V}_0\alpha}{2}\right)+\frac{v^2}{2\psi}+\frac{V_0 }{3} \psi.
\end{equation}

The string tension is given by
\be
\frac{2\kappa ^2}{3}=\frac{7}{6}\bar{V}_0\alpha -\frac{v^2}{2\psi}.
\ee

\begin{figure*}[htbp]
\centering
\includegraphics[width=8.3cm]{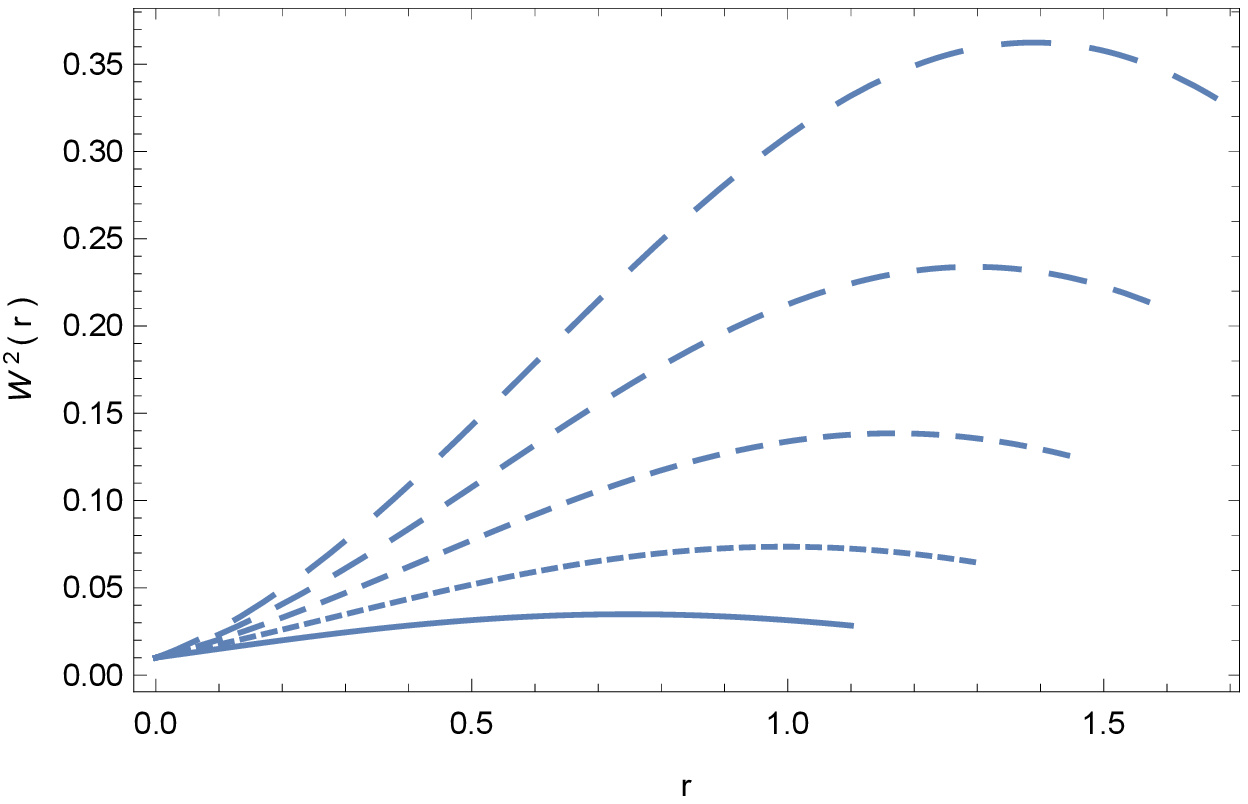}
\includegraphics[width=8.3cm]{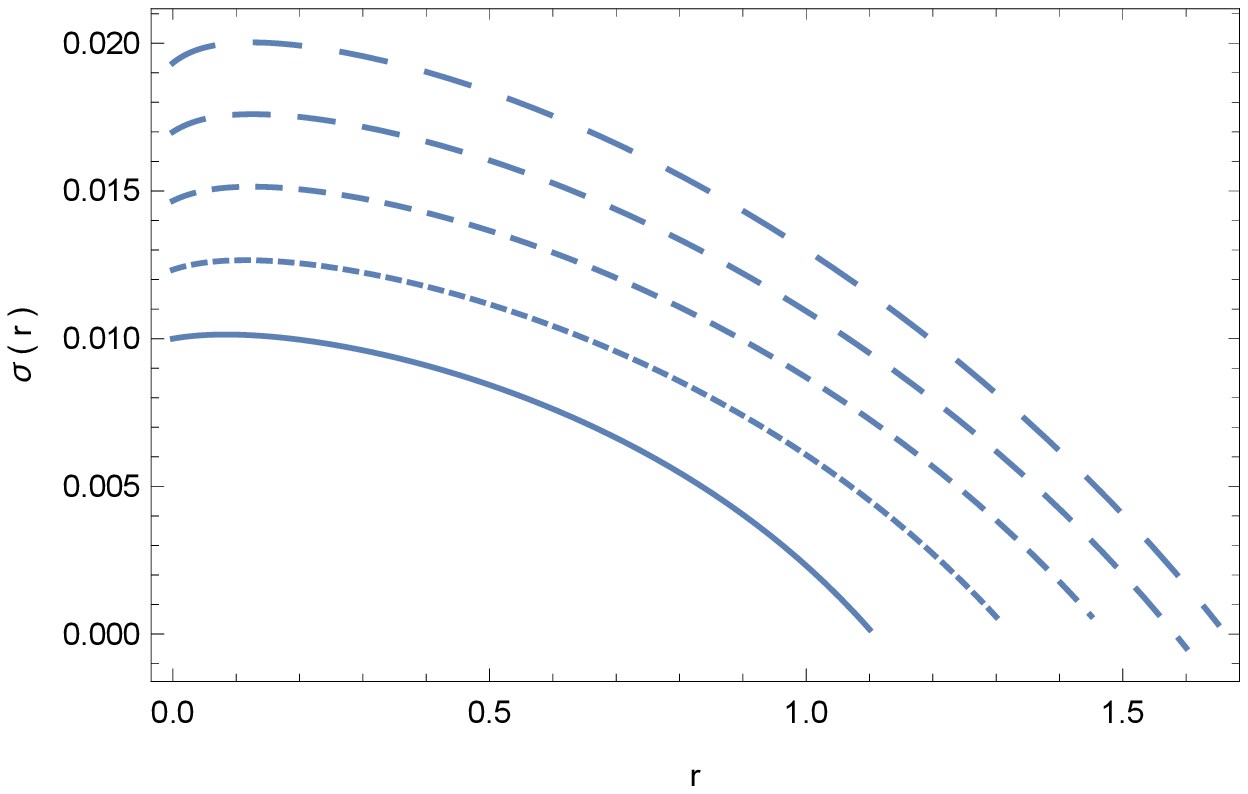}
\caption{Variations of the metric function  $W^2(r)$ (left panel), and of the string tension $\sigma $ (right panel) as a function of $r$ (with all quantities in arbitrary units) for the $\bar{V}(\xi,\psi)=\bar{V}_0\xi^2$ potential, for $\alpha _0=0.010$ (solid curve), $\alpha _0=0.012$ (dotted curve),
 $\alpha _0=0.014$ (short dashed curve), $\alpha _0=0.016$ (dashed curve), and $\alpha _0=0.018$ (long dashed curve), respectively. For $\bar{V}_0$ we have adopted the value $\bar{V}_0=1$, while the boundary conditions used to numerically integrate the field equations are $u_0=-0.001$, $v_0=0.01$, $W(0)=0.10$, and $\psi _0=0.03$, respectively. }
\label{fig1}
\end{figure*}

The system of Eqs.~(\ref{eq1a})-(\ref{eq4a}) must be integrated with the initial conditions $\alpha (0)=\alpha _0$, $\psi(0)=\psi_0$, $W(0)=W_0$, $u(0)=u_0$, and $v(0)=v_0$, respectively. The variations of the metric tensor component $W^2(r)$ and of the string tension $\sigma(r)$ are represented, for different values of $\alpha _0$, and for fixed initial conditions of the other parameters, in Fig.~\ref{fig1}, respectively,

\begin{figure*}[htbp]
\centering
\includegraphics[width=8.3cm]{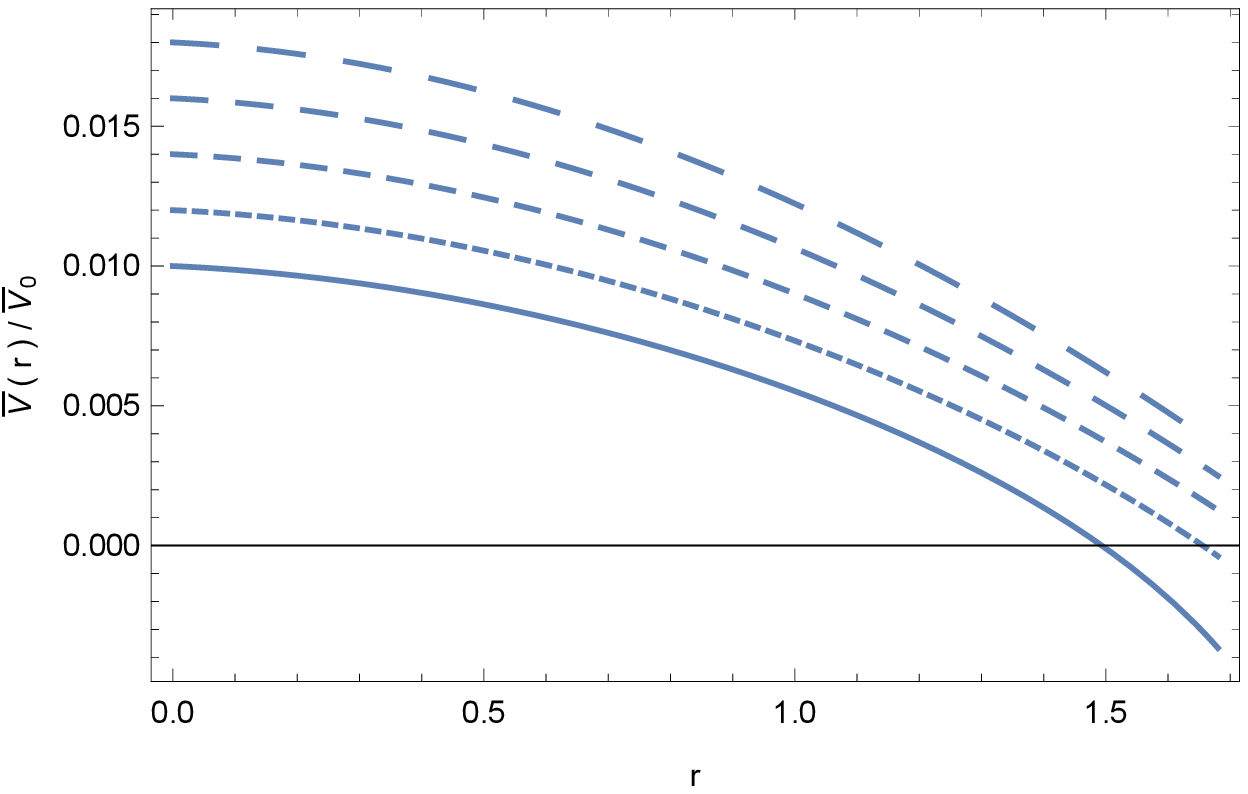}
\includegraphics[width=8.3cm]{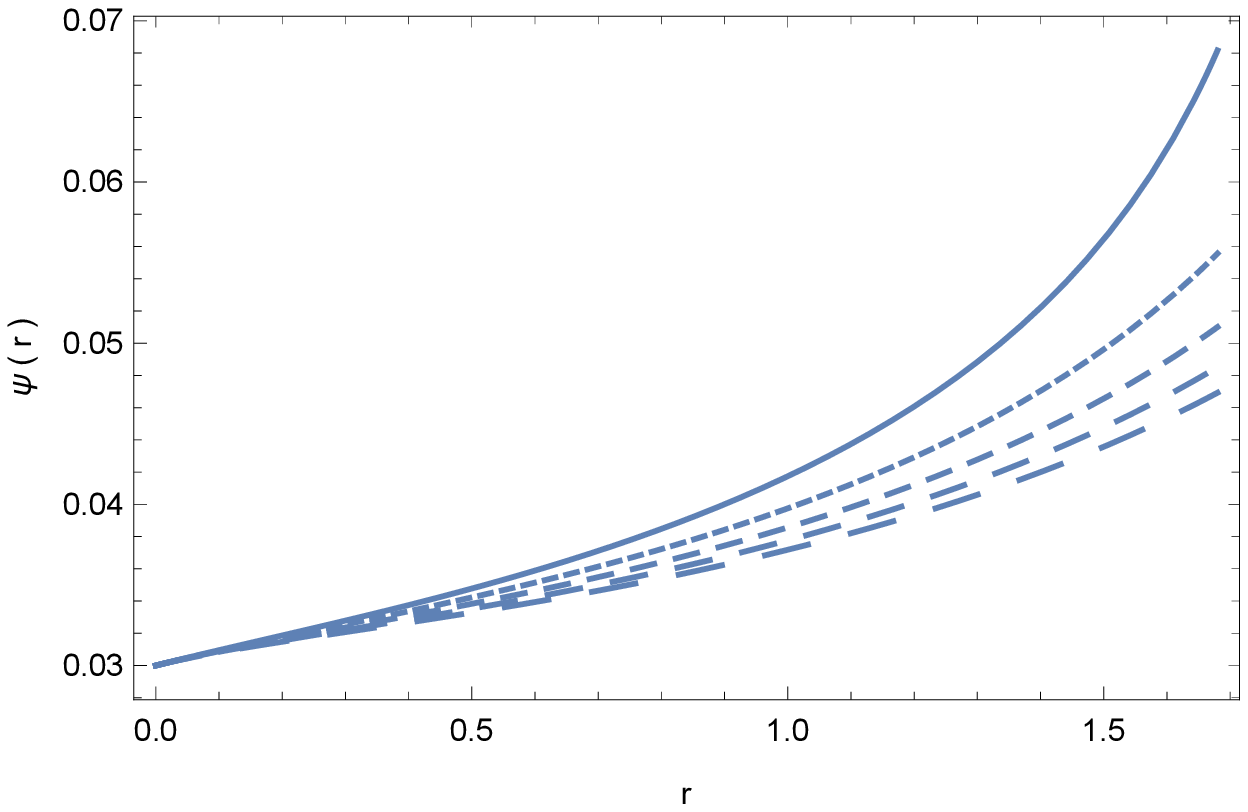}
\caption{Variations of the potential $\bar{V}(\xi,\psi)=\bar{V}_0\xi ^2$ (left panel), and of the function $\psi $ (right panel)  as a function of $r$ (with all quantities in arbitrary units) for the $\bar{V}(\xi)=\bar{V}_0\xi^2$ potential, for $\alpha _0=0.010$ (solid curve), $\alpha _0=0.012$ (dotted curve), $\alpha _0=0.014$ (short dashed curve), $\alpha _0=0.016$ (dashed curve), and $\alpha _0=0.018$ (long dashed curve), respectively. For $\bar{V}_0$ we have adopted the value $\bar{V}_0=1$, while the boundary conditions used to numerically integrate the field equations are $u_0=-0.001$, $v_0=0.01$, $W(0)=0.10$, and $\psi _0=0.03$, respectively. }
\label{fig2}
\end{figure*}

The metric tensor component $W^2$ inside the string takes a finite value for $r=0$, and initially it increases with increasing $r$, reaching a maximum value at $r=r_{max}$. For $r>r_{max}$, $W^2(r)$  becomes a monotonically decreasing function. The behavior of $W^2$ is strongly influenced by the initial conditions, as one can see from its dependence on $\alpha _0$. The string tension $\sigma$ is a monotonically decreasing function of distance, and it identically vanishes at $r=R_s$, with $\sigma \left(R_s\right)=0$. This condition allows to uniquely define $R_s$ as the string radius. The behavior of the string tension is depends strongly on the initial conditions, and this dependence also induces a significant variation of the string radius on the initial values of the string parameters. The variations of the potential $\bar{V}$ and of the function $\psi$ are represented in Fig.~\ref{fig2}.

The potential $\bar{V}(\xi)=\bar{V}_0\xi^2$ is a decreasing function of $r$, having an approximate parabolic dependence on $r$. Inside the string $\xi ^2>0$ for $0\leq r\leq R_s$, which guaranties that the gravitational coupling has the correct sign. The function $\psi$ monotonically increases inside the string, and takes only positive values. The behavior of both functions $V$ and $\psi$ depends strongly on the initial conditions. The string geometric and physical characteristics also depend on the potential parameter $V_0$, but its variation does not change the qualitative behavior of the numerical solution.

\subsubsection{$\bar{V}(\xi,\psi)=\bar{V}_0\xi^4$}

For the $\bar{V}(\xi,\psi)=\bar{V}_0\xi^4$ potential, Eq.~(\ref{Vin1}) takes the form
\begin{equation}
-f\left( R,\mathcal{R}\right) +R \frac{\partial
f\left( R,\mathcal{R}\right) }{\partial R}+\mathcal R\frac{\partial f\left( R,\mathcal{R}\right) }{%
\partial \mathcal{R}}=\bar V_0\left(\frac{\partial
f\left( R,\mathcal{R}\right) }{\partial R}+\frac{\partial f\left( R,\mathcal{R}\right) }{%
\partial \mathcal{R}}\right)^2,
\end{equation}
which yields the following solution to $f\left(R,\mathcal R\right)$
\begin{equation}
f\left(R,\mathcal R\right)=\frac{\left(R+\mathcal R\right)^2}{16 \bar V_0}+g\left(R-\mathcal R\right),
\end{equation}
where $g$ is an arbitrary function.

The structure equations describing the cosmic string configuration in generalized Hybrid Metric-Palatini Gravity take the form
\begin{equation}\label{eq1ab}
\frac{d\alpha }{dr}=u, \qquad \frac{d\psi }{dr}=v,
\end{equation}%
\begin{equation}\label{eq2ab}
\frac{dW}{dr}=\frac{1}{2u}\left( \frac{3v^{2}}{2\psi }-\bar{V}_0\alpha^2\right) W,
\end{equation}%
\begin{equation}\label{eq3ab}
\frac{du}{dr}=-\frac{3v^2}{4\psi}-\frac{\bar{V}_0\alpha ^2}{2},
\end{equation}%
\begin{equation}\label{eq4ab}
\frac{dv}{dr}=-\frac{v}{u}\left(\frac{3v^2}{4\psi}-\frac{\bar{V}_0\alpha ^2}{2}\right)+\frac{v^2}{2\psi}+\frac{2V_0 }{3} \psi \alpha.
\end{equation}

For the string tension we obtain
\be
\frac{2\kappa ^2}{3}=\frac{3}{2}\bar{V}_0\alpha ^2 -\frac{v^2}{2\psi}.
\ee

\begin{figure*}[htbp]
\centering
\includegraphics[width=8.3cm]{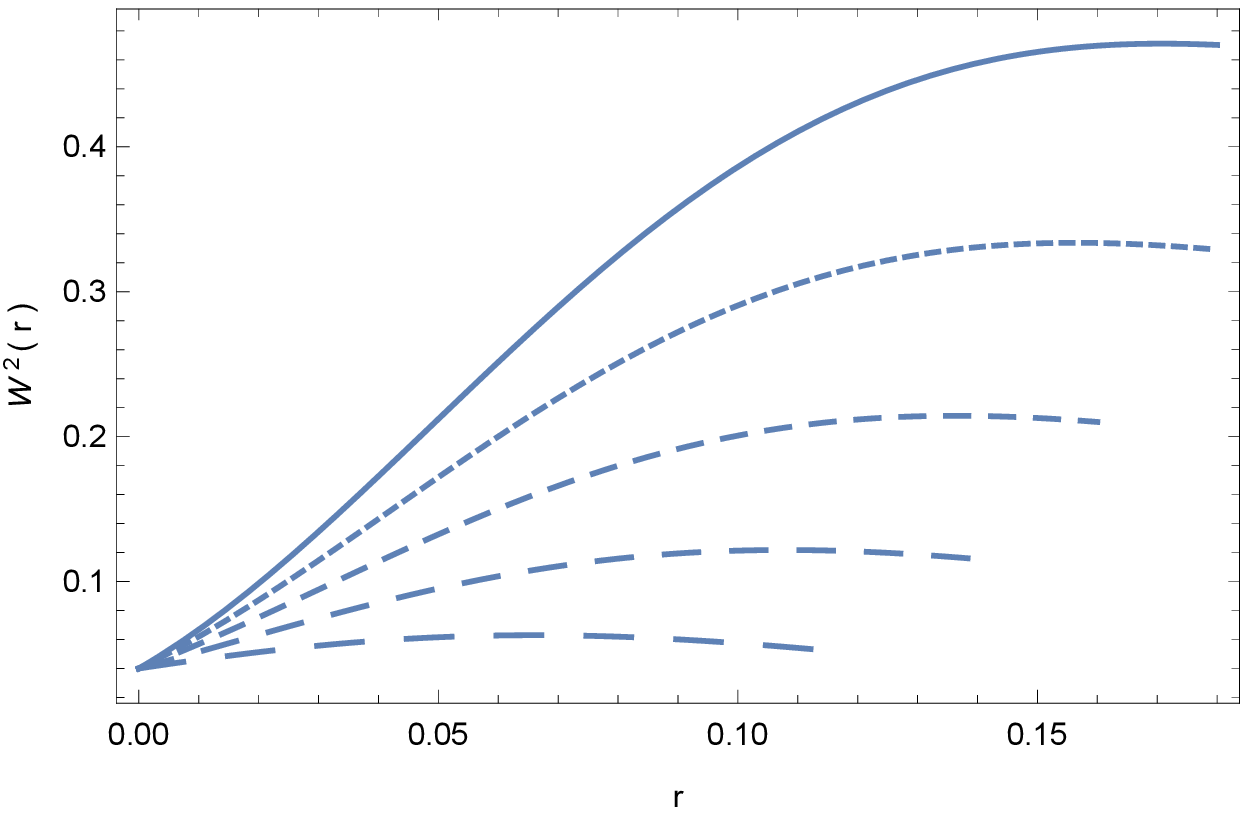}
\includegraphics[width=8.3cm]{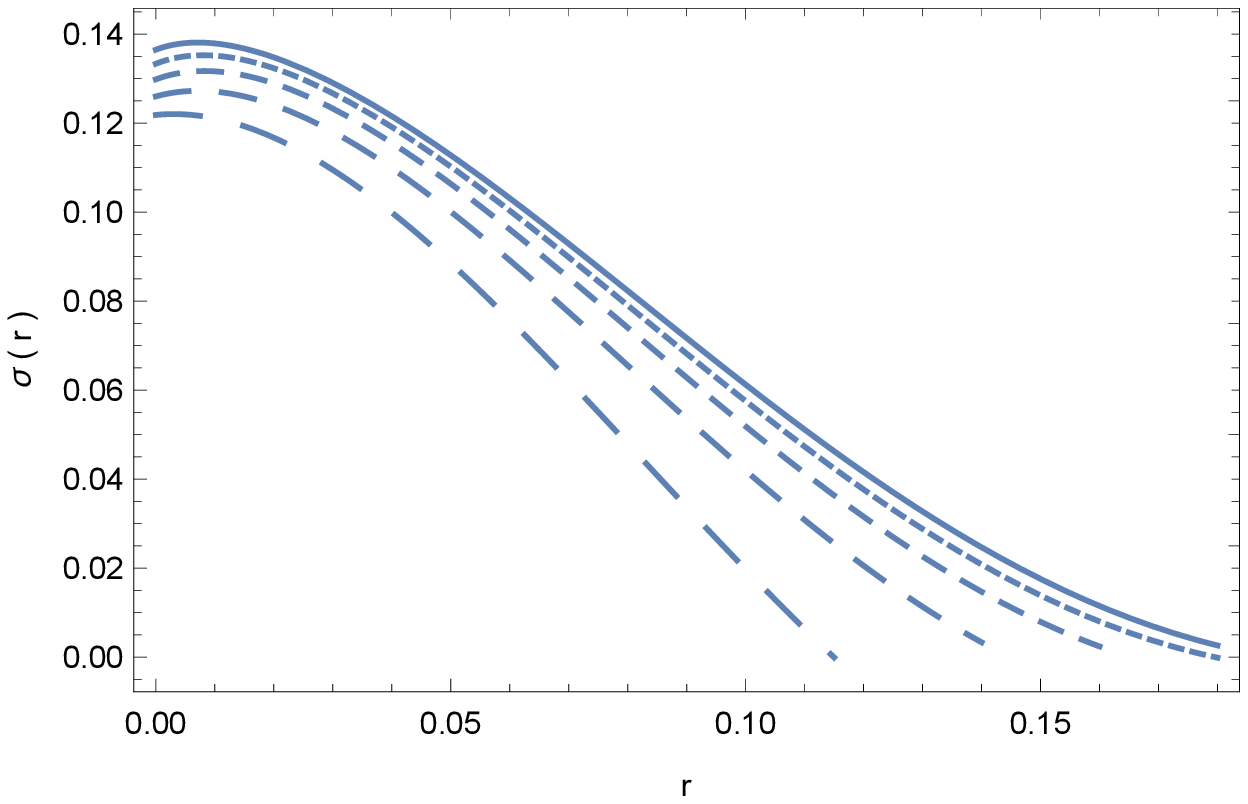}
\caption{Variation of the metric function  $W^2(r)$ (left panel), and of the string tension $\sigma $ (right panel) as a function of $r$ (with all quantities in arbitrary units) for the $\bar{V}(\xi,\psi)=\bar{V}_0\xi^4$ potential, for $v _0=0.00009$ (solid curve), $v _0=0.0001$ (dotted curve),
 $v _0=0.00011$ (short dashed curve), $v _0=0.00012$ (dashed curve), and $v _0=0.00013$ (long dashed curve), respectively. For $\bar{V}_0$ we have adopted the value $\bar{V}_0=10^5$, while the boundary conditions used to numerically integrate the field equations are $u_0=-0.001$, $\alpha_0=0.001$, $W(0)=0.20$, and $\psi _0=0.0000003$, respectively. }
\label{fig1a}
\end{figure*}

The solutions of the system of Eqs.~(\ref{eq1ab})-(\ref{eq4ab}) can be obtained only numerically, and for its integration the initial conditions $\alpha (0)=\alpha _0$, $\psi(0)=\psi_0$, $W(0)=W_0$, $u(0)=u_0$, and $v(0)=v_0$ must be specified. The metric tensor component $W^2(r)$ and the string tension $\sigma(r)$ are represented, for different values of $v _0$, and for fixed initial conditions of the other parameters, in Fig.~\ref{fig1a}, respectively,

\begin{figure*}[htbp]
\centering
\includegraphics[width=8.3cm]{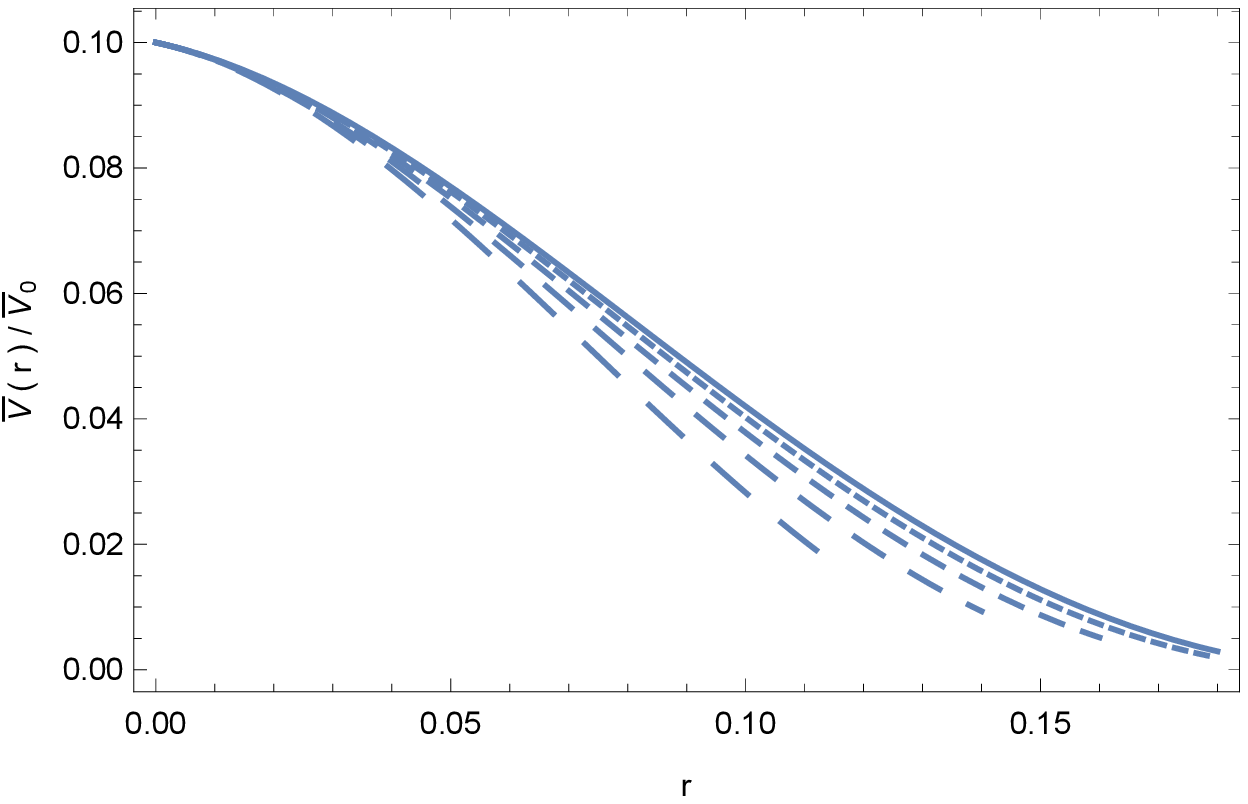}
\includegraphics[width=8.3cm]{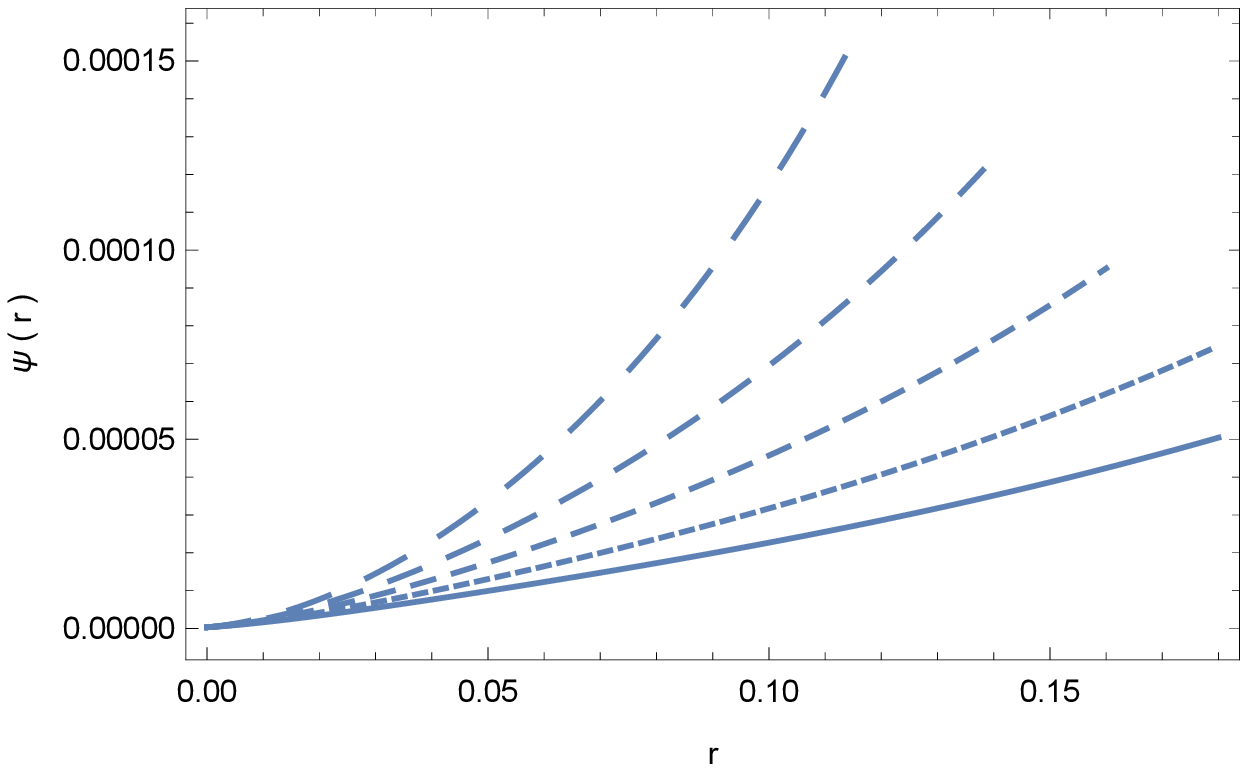}
\caption{Variation of the potential $\bar{V}(\xi,\psi)=\bar{V}_0\xi ^4$ (left panel), and of the function $\psi $ (right panel)  as a function of $r$ (with all quantities in arbitrary units) for the $\bar{V}(\xi)=\bar{V}_0\xi^4$ potential,  for $v _0=0.00009$ (solid curve), $v _0=0.0001$ (dotted curve),
 $v _0=0.00011$ (short dashed curve), $v _0=0.00012$ (dashed curve), and $v _0=0.00013$ (long dashed curve), respectively. For $\bar{V}_0$ we have adopted the value $\bar{V}_0=10^5$, while the boundary conditions used to numerically integrate the field equations are $u_0=-0.001$, $\alpha_0=0.001$, $W(0)=0.20$, and $\psi _0=0.0000003$, respectively. }
\label{fig2a}
\end{figure*}

The qualitative behavior of the metric tensor component $W^2$ inside the string shows some differences as compared to the $\bar{V}(\xi)=\bar{V}_0\xi ^2$ case. While in the quadratic case $W^2$ reaches a maximum inside the string, for the quartic potential the maximal value of the metric tensor is attained on the string surface. The behavior of $W^2$ is strongly dependent on the initial conditions, and significant variations may occur once  $v_0$ is modified. The string tension $\sigma$ monotonically decreases with increasing $r$, and it becomes zero at $r=R_s$, with $\sigma \left(R_s\right)=0$. Hence, we can uniquely define $R_s$ as the string radius. The string tension, as well as the string radius  depend significantly on the initial condition. The variations of the potential $\bar{V}$ and of the function $\psi$ are represented in Fig.~\ref{fig2a}.

The potential $\bar{V}(\xi)=\bar{V}_0\xi^4$ monotonically decreases inside the string, and it becomes zero on the string surface. Similarly to the previous case,  $\xi ^2>0$ for $0\leq r\leq R_s$, and thus the gravitational coupling has the correct sign. The function $\psi$ monotonically increases, and takes only positive values. The behavior of both functions $V$ and $\psi$ is significantly dependent on the initial conditions, and on the potential parameter $V_0$. However, major changes in the numerical value of $V_0$ do not affect significantly the qualitative behavior of the numerical string solution for the quartic potential of the generalized Hybrid Metric-Palatini Gravity theory.

\subsection{$\bar{V}\left(\xi,\psi\right)=\bar{V}_0\psi ^2$}

Next we assume that the potential $\bar{V}$ is independent on the field $\xi$, and has the form $\bar{V}=\bar{V}_0\psi ^2$, with $\bar{V}_0$ a constant. For this choice of the potential, Eq.~(\ref{Vin1}) becomes
\begin{equation}
-f\left( R,\mathcal{R}\right) +R \frac{\partial
f\left( R,\mathcal{R}\right) }{\partial R}+\mathcal R\frac{\partial f\left( R,\mathcal{R}\right) }{%
\partial \mathcal{R}}=\bar V_0 \frac{\partial f\left( R,\mathcal{R}\right) }{%
\partial \mathcal{R}}
\end{equation}
and a particular solution for the function $f\left(R,\mathcal R\right)$ is
\begin{equation}
f\left(R,\mathcal R\right)=\frac{\left(\mathcal R+c_1R\right)^2}{4\bar V_0\left(1+c_2\mathcal R\right)}+c_3 R,
\end{equation}
where the $c_i$ are constants.

The system of field equations describing the cosmic string behavior in the Generalized Hybrid Metric Palatini gravity take the form
\begin{equation}\label{eq1b}
\frac{d\alpha }{dr}=u, \qquad \frac{d\psi }{dr}=v,
\end{equation}%
\begin{equation}\label{eq2b}
\frac{dW}{dr}=\frac{1}{2u}\left( \frac{3v^{2}}{2\psi }-\bar{V}_0\psi ^2\right) W,
\end{equation}%
\begin{equation}\label{eq3b}
\frac{du}{dr}=-\frac{3v^2}{4\psi}-\frac{\bar{V}_0\psi^2}{2},
\end{equation}%
\begin{equation}\label{eq4b}
\frac{dv}{dr}=-\frac{v}{u}\left(\frac{3v^2}{4\psi}-\frac{\bar{V}_0\alpha}{2}\right)+\frac{v^2}{2\psi}+\frac{2V_0 }{3} \psi^2.
\end{equation}

The string tension can be obtained as
\be
\frac{2\kappa ^2}{3}\sigma =\frac{5}{6}\bar{V}_0\psi ^2-\frac{v^2}{2\psi}.
\ee

\begin{figure*}[htbp]
\centering
\includegraphics[width=8.3cm]{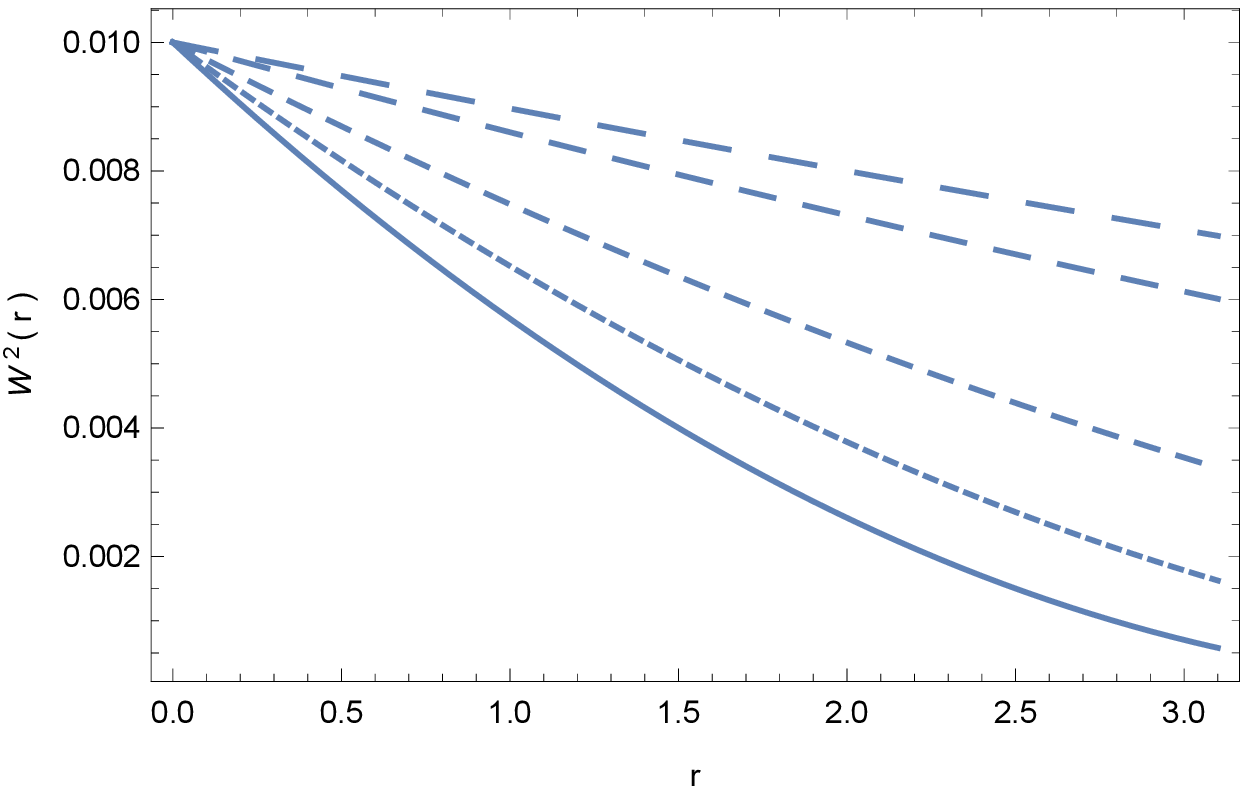}
\includegraphics[width=8.3cm]{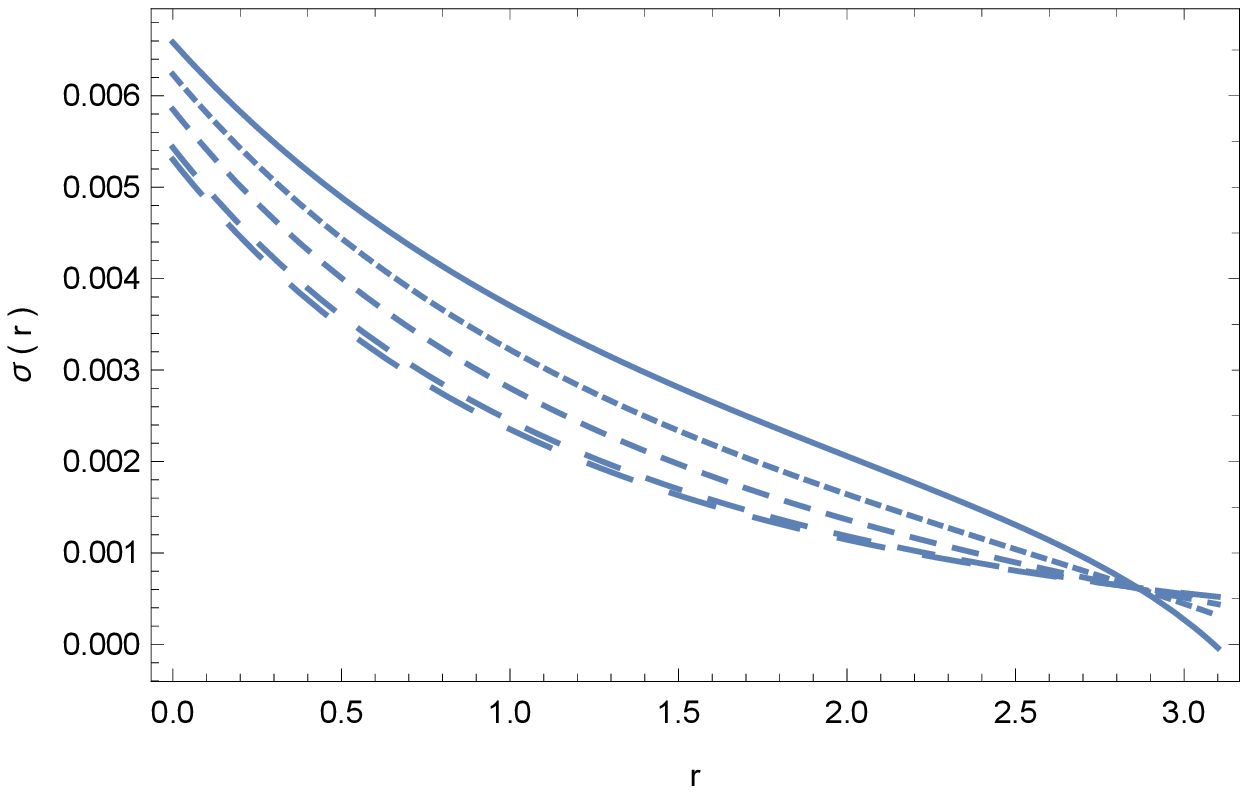}
\caption{Variations of the metric function  $W^2(r)$ (left panel), and of the string tension $\sigma (r) $ (right panel) as a function of $r$ (with all quantities in arbitrary units) for the $\bar{V}(\xi,\psi)=\bar{V}_0\psi^2$ potential, for $v _0=-0.01$ (solid curve), $v _0=-0.011$ (dotted curve),
 $v _0=-0.012$ (short dashed curve), $v _0=-0.013$ (dashed curve), and $v _0=-0.0133$ (long dashed curve), respectively. For $\bar{V}_0$ we have adopted the value $\bar{V}_0=11$, while the boundary conditions used to numerically integrate the field equations are $u_0=-0.001$, $\alpha_0=0.10$, $W(0)=0.10$, and $\psi _0=0.03$, respectively. }
\label{fig3}
\end{figure*}

The variations of the metric function $W^2$ and of the string tension are represented in Fig.~\ref{fig3}. Both $W^2$ and $\sigma$ are monotonically decreasing functions, with $\sigma (r)$ vanishing for a finite value of $r=R_s$, which represents the radius of the string. To obtain the plots we have varied the initial condition for $\psi '(0)=v_0$. There is a significant impact on the numerical value of $v_0$ on the string properties. However, even that for small $r$ the effect on the string tension is rather large, the string radius is less affected by the variation of this initial condition.

\begin{figure*}[htbp]
\centering
\includegraphics[width=8.3cm]{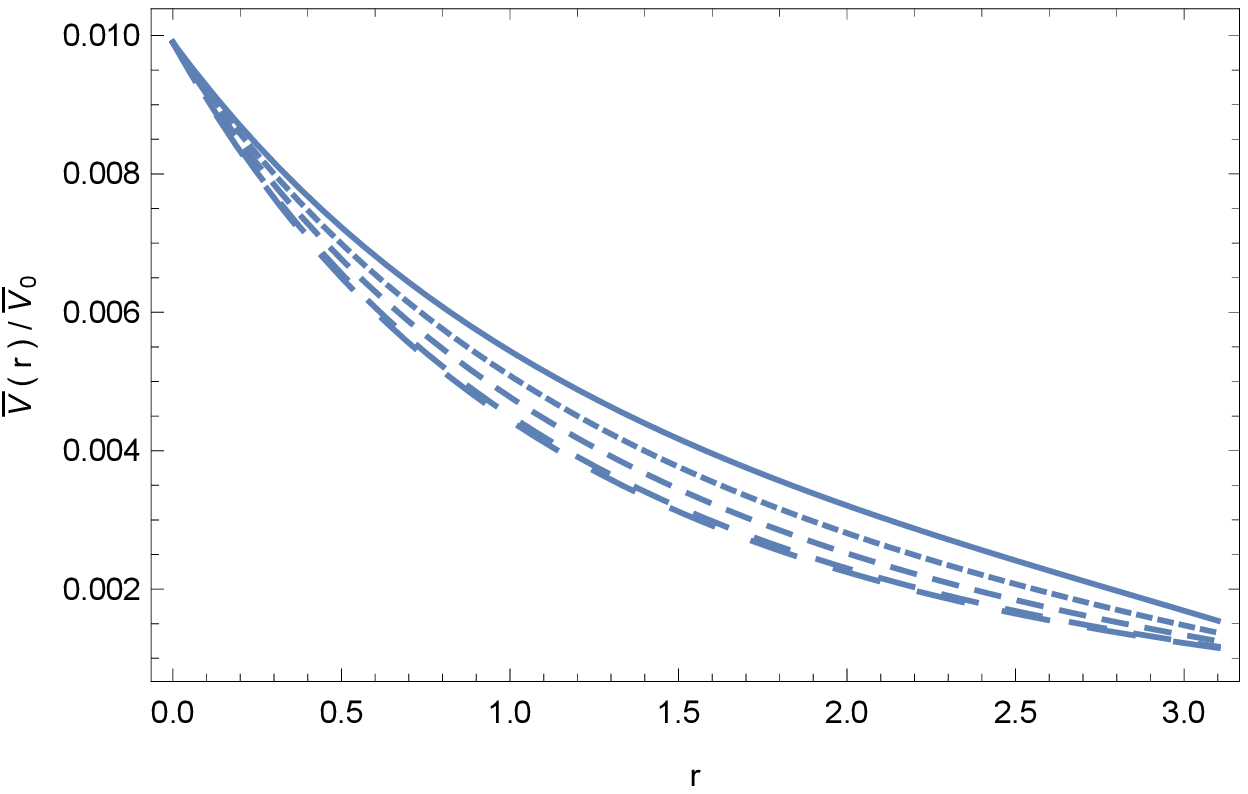}
\includegraphics[width=8.3cm]{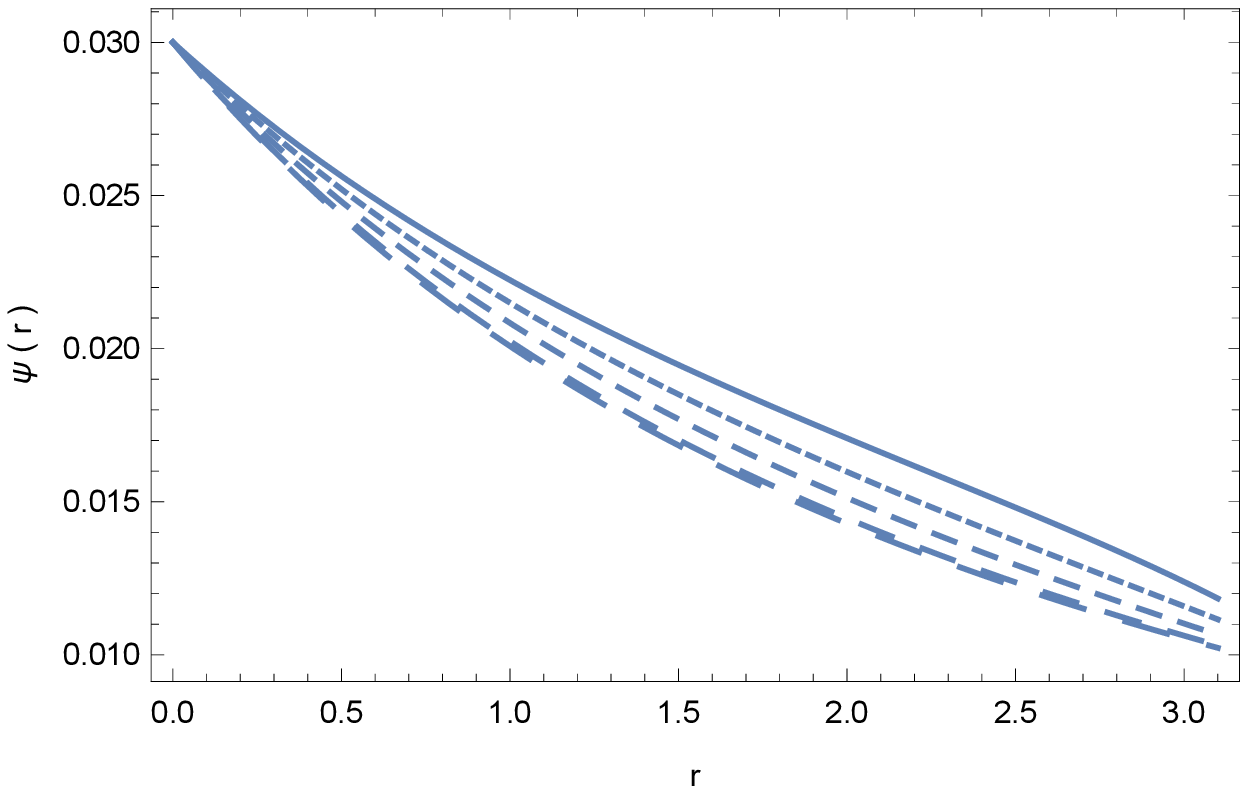}
\caption{Variations of the potential  $\bar{V}(r)$ (left panel), and of the function $\psi (r) $ (right panel) as a function of $r$ (with all quantities in arbitrary units) for the $\bar{V}(\xi,\psi)=\bar{V}_0\psi^2$ potential, for $v _0=-0.01$ (solid curve), $v _0=-0.011$ (dotted curve),
 $v _0=-0.012$ (short dashed curve), $v _0=-0.013$ (dashed curve), and $v _0=-0.0133$ (long dashed curve), respectively. For $\bar{V}_0$ we have adopted the value $\bar{V}_0=11$, while the boundary conditions used to numerically integrate the field equations are $u_0=-0.001$, $\alpha_0=0.10$, $W(0)=0.10$, and $\psi _0=0.03$, respectively. }
\label{fig4}
\end{figure*}

The variations of the potential $\bar{V}$ and of the function $\psi$ are represented in Fig.~\ref{fig4}. Both quantities monotonically decrease radially, and their behavior depends on the adopted initial conditions.

\begin{figure}[htbp]
\centering
\includegraphics[width=8.3cm]{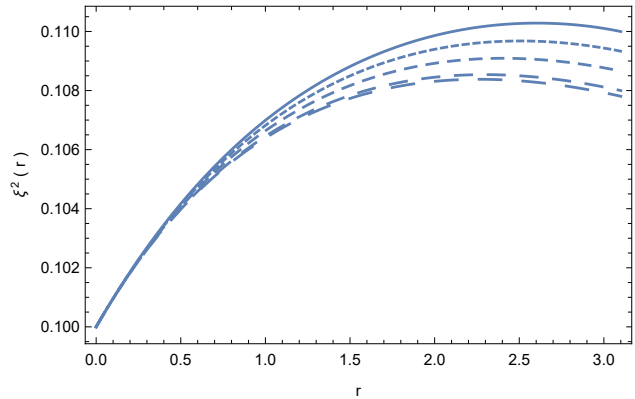}
\caption{Variation of $\xi^2$ as a function of $r$ (with all quantities in arbitrary units) for the $\bar{V}(\xi,\psi)=\bar{V}_0\psi^2$ potential, for $v _0=-0.01$ (solid curve), $v _0=-0.011$ (dotted curve),
 $v _0=-0.012$ (short dashed curve), $v _0=-0.013$ (dashed curve), and $v _0=-0.0133$ (long dashed curve), respectively. For $\bar{V}_0$ we have adopted the value $\bar{V}_0=11$, while the boundary conditions used to numerically integrate the field equations are $u_0=-0.001$, $\alpha_0=0.10$, $W(0)=0.10$, and $\psi _0=0.03$, respectively. }
\label{fig5}
\end{figure}

The variation of $\xi ^2(r)$ is represented in Fig.~\ref{fig5}. $\xi ^2$ is positive in the range $0\leq r\leq R_s$, thus ensuring the physical nature of the gravitational coupling.   $\xi ^2$ is a monotonically increasing function of $r$, and at large values of the radial coordinate, near the string boundary, its variation depends on the initial condition for $\psi '(0)$.

The variation of the potential parameter $\bar{V}_0$ does not change the qualitative behavior of the solution, even that it has an important effect on the numerical characteristics of the string.

\subsection{$\bar{V}\left(\xi,\psi\right)=\bar{V}_0\xi ^2\psi ^2$}

Next we will consider string type solutions in the Generalized Hybrid Metric Palatini Gravity under the assumption that the potential $\bar{V}$ is given by $\bar{V}=\bar{V}_0\xi ^2\psi ^2=\bar{V_0}\alpha \psi ^2$, with $\bar{V}_0$ constant. Equation (\ref{Vin1}) then becomes
\begin{equation}
-f\left( R,\mathcal{R}\right) +R \frac{\partial
f\left( R,\mathcal{R}\right) }{\partial R}+\mathcal R\frac{\partial f\left( R,\mathcal{R}\right) }{%
\partial \mathcal{R}}=\bar V_0 \frac{\partial f\left( R,\mathcal{R}\right) }{%
\partial \mathcal{R}}\left(\frac{\partial f\left( R,\mathcal{R}\right) }{%
\partial \mathcal{R}}+\frac{\partial f\left( R,\mathcal{R}\right) }{%
\partial R}\right)
\end{equation}
and a particular solution for the function $f\left(R,\mathcal R\right)$ is
\begin{equation}
f\left(R,\mathcal R\right)=\sqrt{\frac{R}{\bar V_0}}\left(R-\mathcal R\right).
\end{equation}

For this potential the field equations describing the string-like structure take the form
\begin{equation}\label{eq1c}
\frac{d\alpha }{dr}=u, \qquad \frac{d\psi }{dr}=v,
\end{equation}%
\begin{equation}\label{eq2c}
\frac{dW}{dr}=\frac{1}{2u}\left( \frac{3v^{2}}{2\psi }-\bar{V}_0\alpha \psi ^2\right) W,
\end{equation}%
\begin{equation}\label{eq3c}
\frac{du}{dr}=-\frac{3v^2}{4\psi}-\frac{\bar{V}_0\alpha \psi^2}{2},
\end{equation}%
\begin{equation}\label{eq4c}
\frac{dv}{dr}=-\frac{v}{u}\left(\frac{3v^2}{4\psi}-\frac{\bar{V}_0\alpha}{2}\right)+\frac{v^2}{2\psi}+\frac{2V_0 }{3} \psi^2\left(\alpha +\frac{\psi}{2}\right).
\end{equation}

For this model the string tension is given by
\be
\frac{2\kappa ^2}{3}\sigma=\frac{7}{6}\bar{V}_0\alpha \psi^2-\frac{v^2}{2\psi}.
\ee

\begin{figure*}[htbp]
\centering
\includegraphics[width=8.3cm]{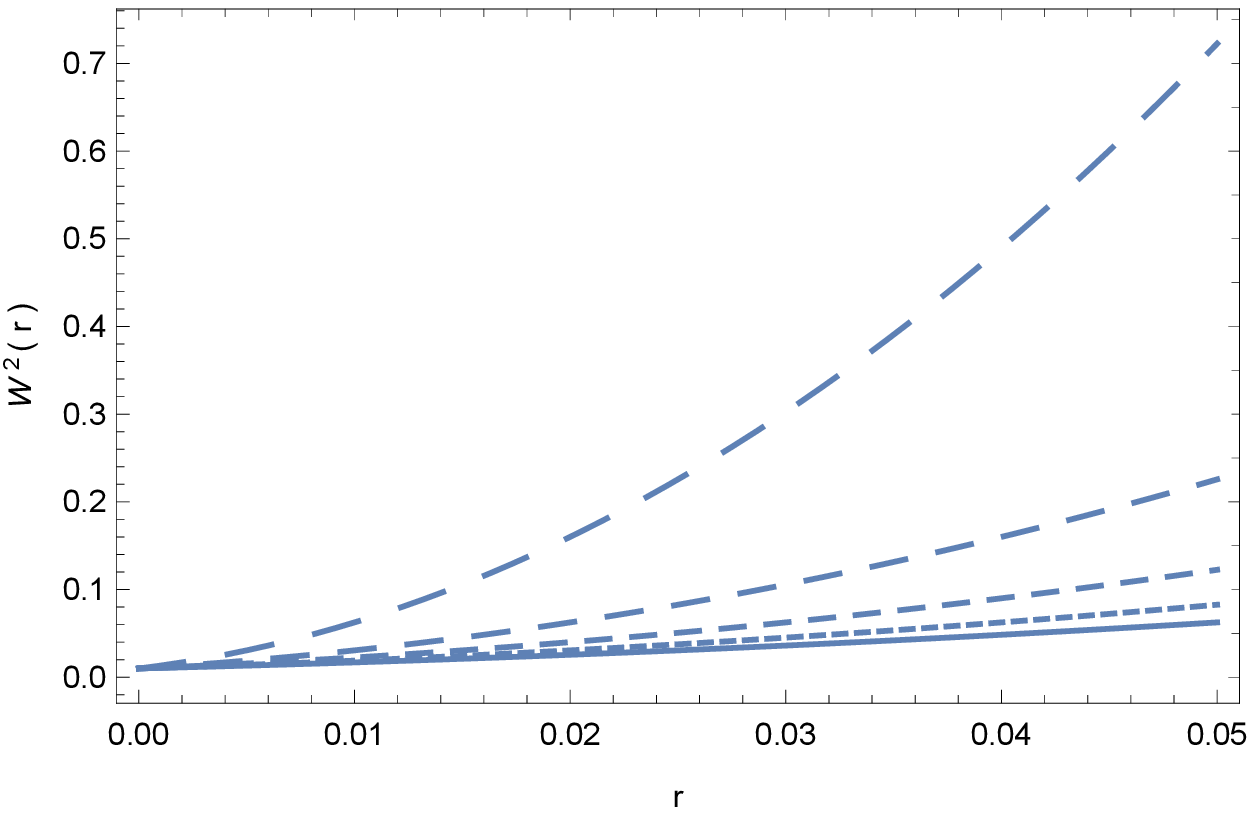}
\includegraphics[width=8.3cm]{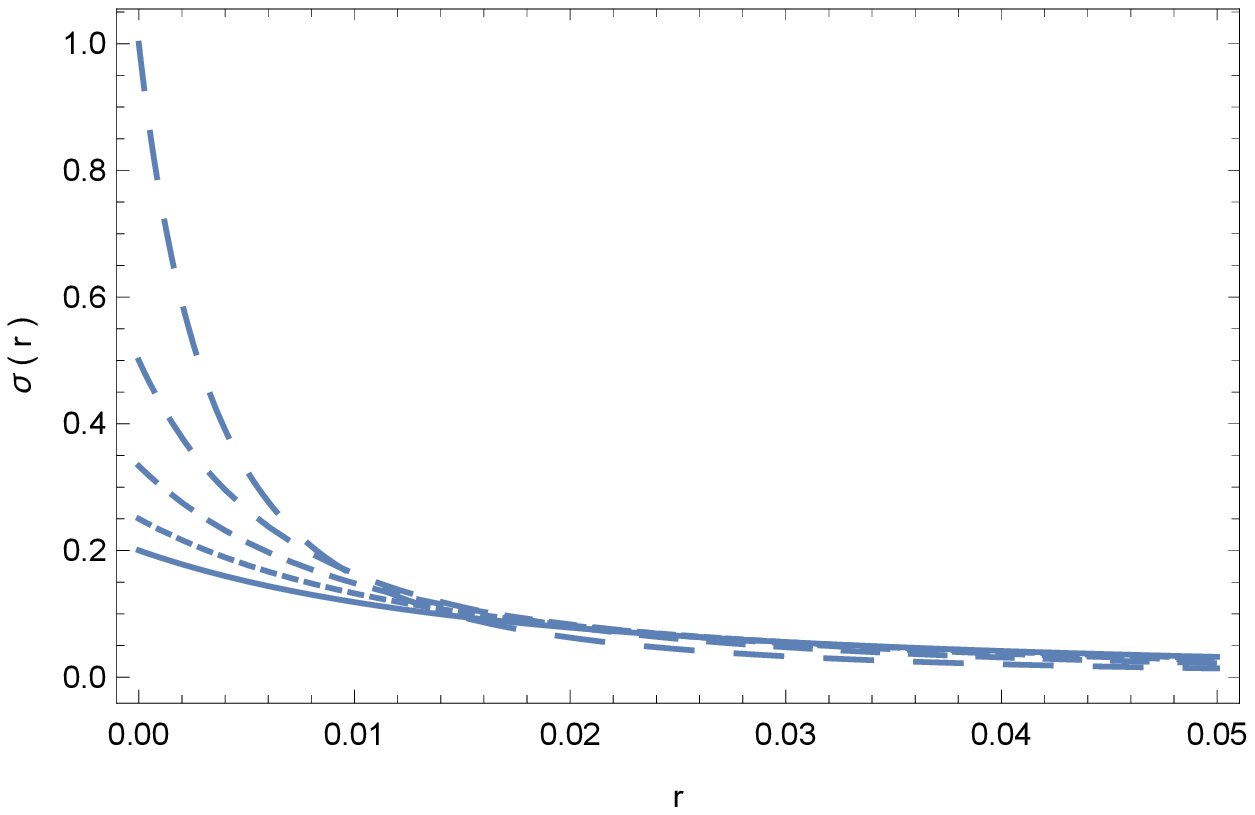}
\caption{Variations of the metric function  $W^2(r)$ (left panel), and of the string tension $\sigma (r) $ (right panel) as a function of $r$ (with all quantities in arbitrary units) for the $\bar{V}(\xi,\psi)=\bar{V}_0\xi ^2\psi^2$ potential, for $\psi _0=-0.025$ (solid curve), $\psi _0=-0.020$ (dotted curve),
 $\psi _0=-0.015$ (short dashed curve), $\psi _0=-0.01$ (dashed curve), and $\psi _0=-0.005$ (long dashed curve), respectively. For $\bar{V}_0$ we have adopted the value $\bar{V}_0=10$, while the boundary conditions used to numerically integrate the field equations are $u_0=-0.01$, $\alpha_0=0.025$, $W(0)=0.10$, and $v_0=0.10$, respectively. }
\label{fig6}
\end{figure*}

The metric function $W^2$ and the string tension $\sigma$ are depicted in Fig.~\ref{fig6}, for a varying initial condition $\psi '(0)=\psi_0$, while all the other initial conditions are fixed. In this case the radial metric function is an increasing function of the radial coordinate $r$, and its rate of increase is strongly dependent on the variations in the numerical values of $\psi _0$. Similarly to the previous cases, the string tension is a monotonically decreasing function of $r$, and it vanishes at a finite value of $r$, $r=R_s$, which uniquely defines the string radius. The string radius is weakly dependent on the variation of $\psi _0$, however, significant variations in $\sigma$ do appear for small values of $r$.

\begin{figure*}[htbp]
\centering
\includegraphics[width=8.3cm]{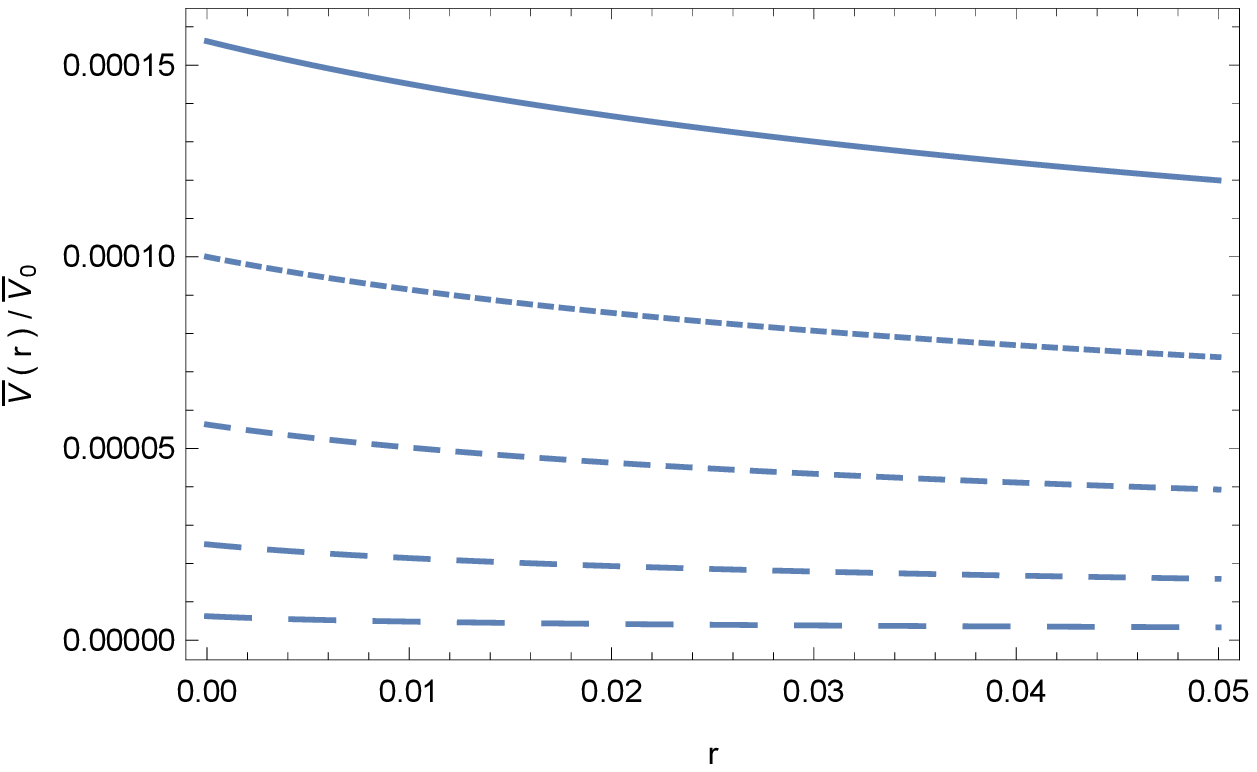}
\includegraphics[width=8.3cm]{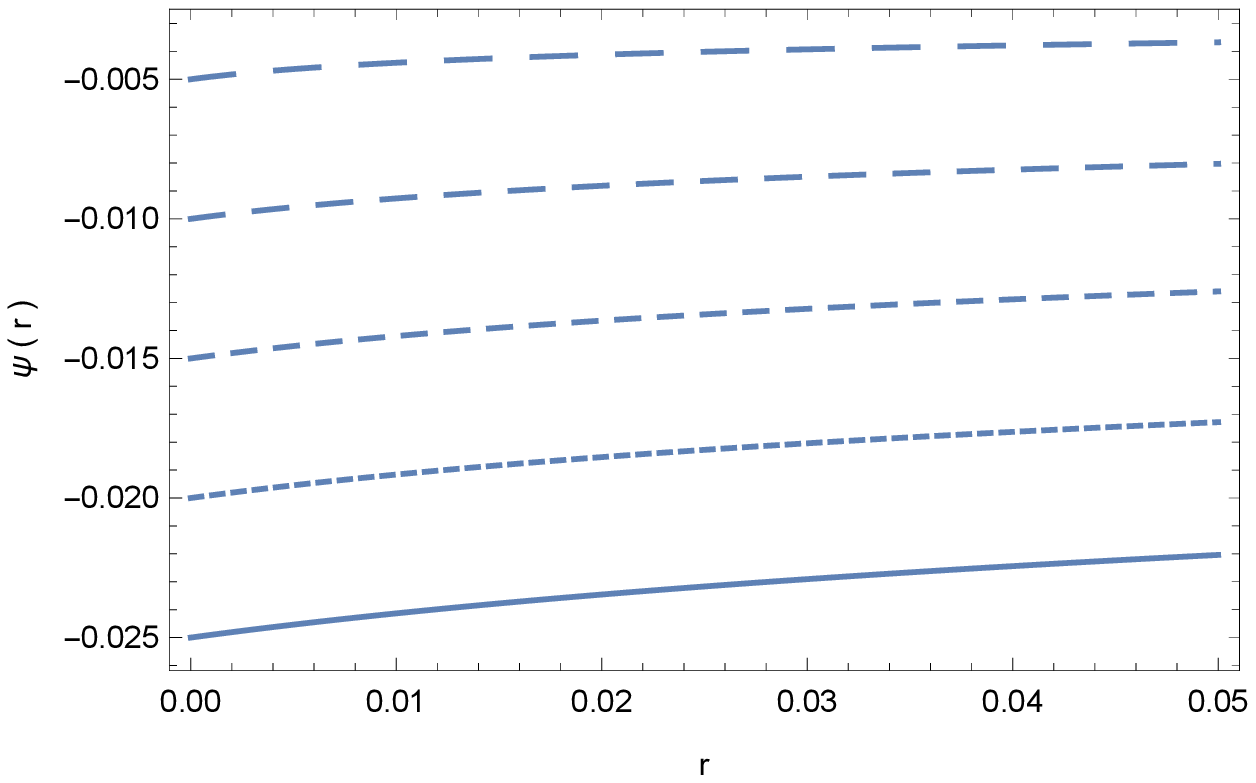}
\caption{Variations of the potential  $\bar{V}\left(\xi,\psi\right)=\bar{V}_0\xi ^2\psi ^2$ (left panel), and of the function $\psi $ (right panel) as a function of $r$ (with all quantities in arbitrary units) for the $\bar{V}(\xi)=\bar{V}_0\xi ^2\psi^2$ potential, for $\psi _0=-0.025$ (solid curve), $\psi _0=-0.020$ (dotted curve),
 $\psi _0=-0.015$ (short dashed curve), $\psi _0=-0.01$ (dashed curve), and $\psi _0=-0.005$ (long dashed curve), respectively. For $\bar{V}_0$ we have adopted the value $\bar{V}_0=10$, while the boundary conditions used to numerically integrate the field equations are $u_0=-0.01$, $\alpha_0=0.025$, $W(0)=0.10$, and $v_0=0.10$, respectively. }
\label{fig7}
\end{figure*}

The variations of the potential and of the function $\psi$ are represented in Fig.~\ref{fig7}. $\bar{V}$ is a slowly decreasing positive function of $r$, strongly dependent on the initial condition for $\psi'$. The function $\psi$ takes negative values, and show a strong dependence on $\psi _0$.

\begin{figure}[htbp]
\centering
\includegraphics[width=8.3cm]{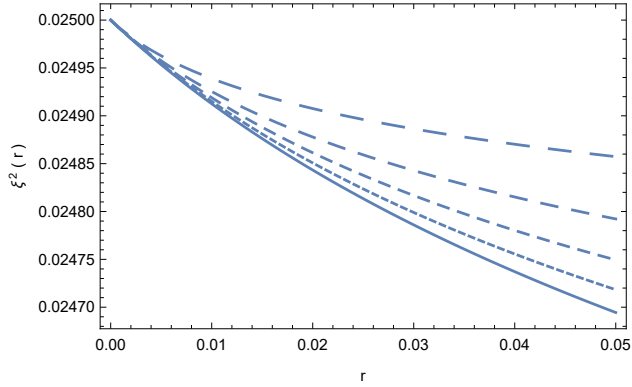}
\caption{Variation of $\xi^2$ as a function of $r$ (with all quantities in arbitrary units) for the $\bar{V}(\xi,\psi)=\bar{V}_0\xi^2 \psi^2$ potential, for $\psi _0=-0.025$ (solid curve), $\psi _0=-0.020$ (dotted curve),
 $\psi _0=-0.015$ (short dashed curve), $\psi _0=-0.01$ (dashed curve), and $\psi _0=-0.005$ (long dashed curve), respectively. For $\bar{V}_0$ we have adopted the value $\bar{V}_0=10$, while the boundary conditions used to numerically integrate the field equations are $u_0=-0.01$, $\alpha_0=0.025$, $W(0)=0.10$, and $v_0=0.10$, respectively. }
\label{fig8}
\end{figure}

The behavior of the function $\xi ^2(r)$ is depicted in Fig.~\ref{fig8}. $\xi ^2$ is positive for $r\in \left[0,R_s\right]$, and thus the physical nature of the gravitational coupling in the present model is guaranteed.   $\xi ^2$ is a monotonically decreasing function of $r$, and  its variation depends significantly on the numerical values of the initial conditions for $\psi '(0)$.

\subsection{$\bar{V}(\xi,\psi)=a\xi ^2+b\psi ^2$}

Finally,  consider string type solutions in the Generalized Hybrid Metric Palatini Gravity by assuming that the potential $\bar{V}$ is given by the quadratic expression  $\bar{V}=a\xi ^2+b\psi ^2=a\alpha +b\psi ^2$, with $a$, $b$ constants. In this case, Eq.~(\ref{Vin1}) takes the form
\begin{equation}
-f\left( R,\mathcal{R}\right) +R \frac{\partial
f\left( R,\mathcal{R}\right) }{\partial R}+\mathcal R\frac{\partial f\left( R,\mathcal{R}\right) }{%
\partial \mathcal{R}}=a\frac{\partial f\left( R,\mathcal{R}\right) }{%
\partial \mathcal{R}}+b\left(\frac{\partial f\left( R,\mathcal{R}\right) }{%
\partial \mathcal{R}}+\frac{\partial f\left( R,\mathcal{R}\right) }{%
\partial R}\right)
\end{equation}
and a particular solution for the function $f\left(R,\mathcal R\right)$ is
\begin{equation}
f\left(R,\mathcal R\right)=\frac{1}{4b}\left[a\left(a-R\right)+\left(R-\mathcal R\right)^2\right]+c\left(R-a\right),
\end{equation}
where $c$ is a constant.

The field equations describing the string-like structure take the form
\begin{equation}\label{eq1c}
\frac{d\alpha }{dr}=u, \qquad \frac{d\psi }{dr}=v,
\end{equation}%
\begin{equation}\label{eq2c}
\frac{dW}{dr}=\frac{1}{2u}\left( \frac{3v^{2}}{2\psi }-a\alpha-b\psi^2\right) W,
\end{equation}%
\begin{equation}\label{eq3c}
\frac{du}{dr}=-\frac{3v^2}{4\psi}-\frac{1}{2}\left(a\alpha +b\psi ^2\right),
\end{equation}%
\begin{equation}\label{eq4c}
\frac{dv}{dr}=-\frac{v}{u}\left(\frac{3v^2}{4\psi}-\frac{\bar{V}_0\alpha}{2}\right)+\frac{v^2}{2\psi}+\frac{\psi}{3} \left(a+2b\psi\right).
\end{equation}

For this model the string tension is given by
\be
\frac{2\kappa ^2}{3}\sigma=\frac{7}{6}a\alpha +\frac{5}{6}b\psi^2-\frac{v^2}{2\psi}.
\ee

In the following we will consider two class of models described by this potential, obtained by varying the potential parameters $(a,b)$ for fixed initial conditions, and models in which the potential parameters are fixed, while the initial conditions for $r=0$ are slightly modified.

\subsubsection{Varying the potential parameters}

\begin{figure*}[htbp]
\centering
\includegraphics[width=8.3cm]{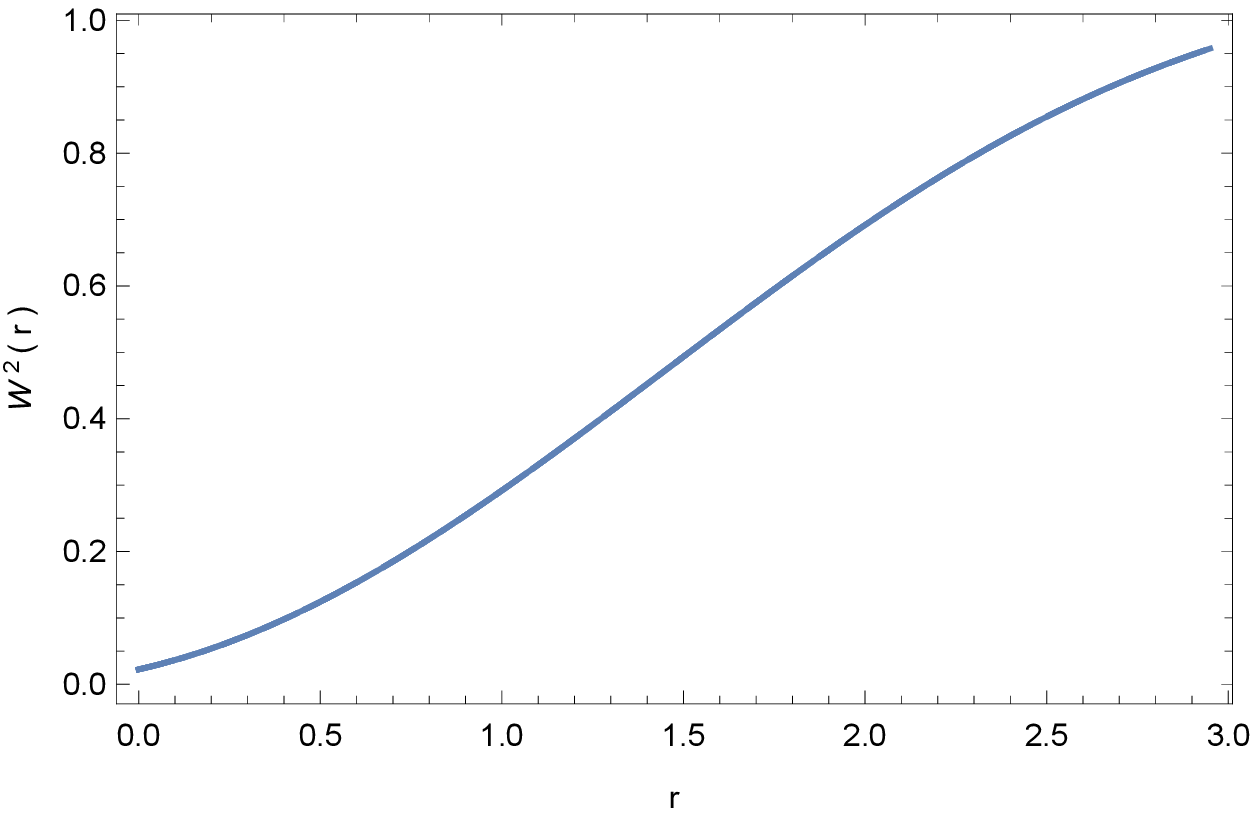}
\includegraphics[width=8.3cm]{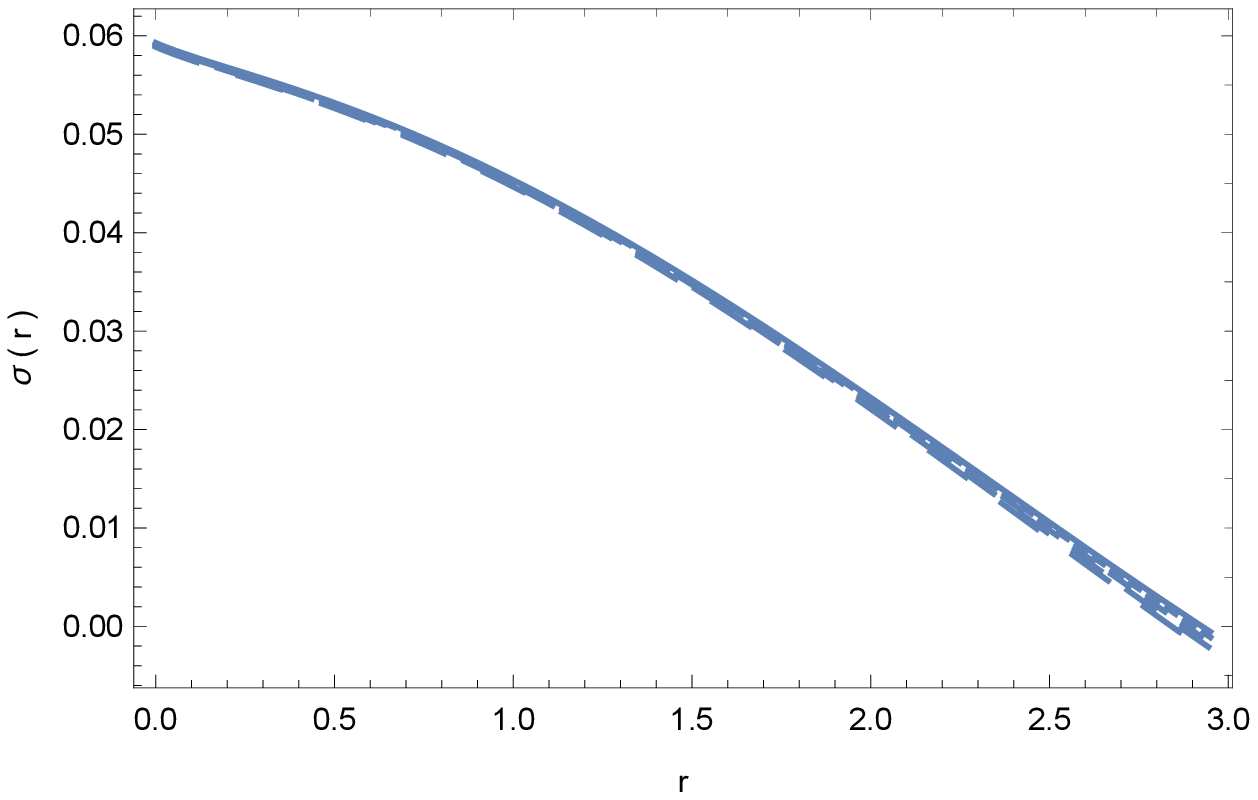}
\caption{Variations of the metric function  $W^2(r)$ (left panel), and of the string tension $\sigma (r) $ (right panel) as a function of $r$ (with all quantities in arbitrary units) for the $\bar{V}(\xi,\psi)=a\xi ^2+b\psi^2$ potential, for $a=0.5$ and $b=-2$ (solid curve), $b=-2.2$ (dotted curve),
 $b=-2.4$ (short dashed curve), $b=-2.6$ (dashed curve), and $b=-2.8$ (long dashed curve), respectively. The boundary conditions used to numerically integrate the field equations are $u_0=-0.01$, $\alpha_0=0.10$, $W(0)=0.15$, and $v_0=-0.01$, respectively. }
\label{fig9}
\end{figure*}

As a first example of string solutions with the quadratic potential in $\xi $ and $\psi$ we consider a configuration with fixed initial conditions but for different values of $a$ and $b$. For the sake of concreteness we fix $a$ and we vary $b$. The variation of the metric function $W^2$ and of the string tension are represented in Fig.~\ref{fig9} for different values of $b$. $W^2(r)$ is a monotonically increasing function of $r$, reaching its maximum value on the string vacuum boundary. The string tension monotonically decreases from its value at $r=0$ to zero, with the corresponding value of $r$ uniquely determining the string boundary. Both the variations of $W^2$ and $\sigma $ are basically independent of the variations of the potential parameters.

\begin{figure*}[htbp]
\centering
\includegraphics[width=8.3cm]{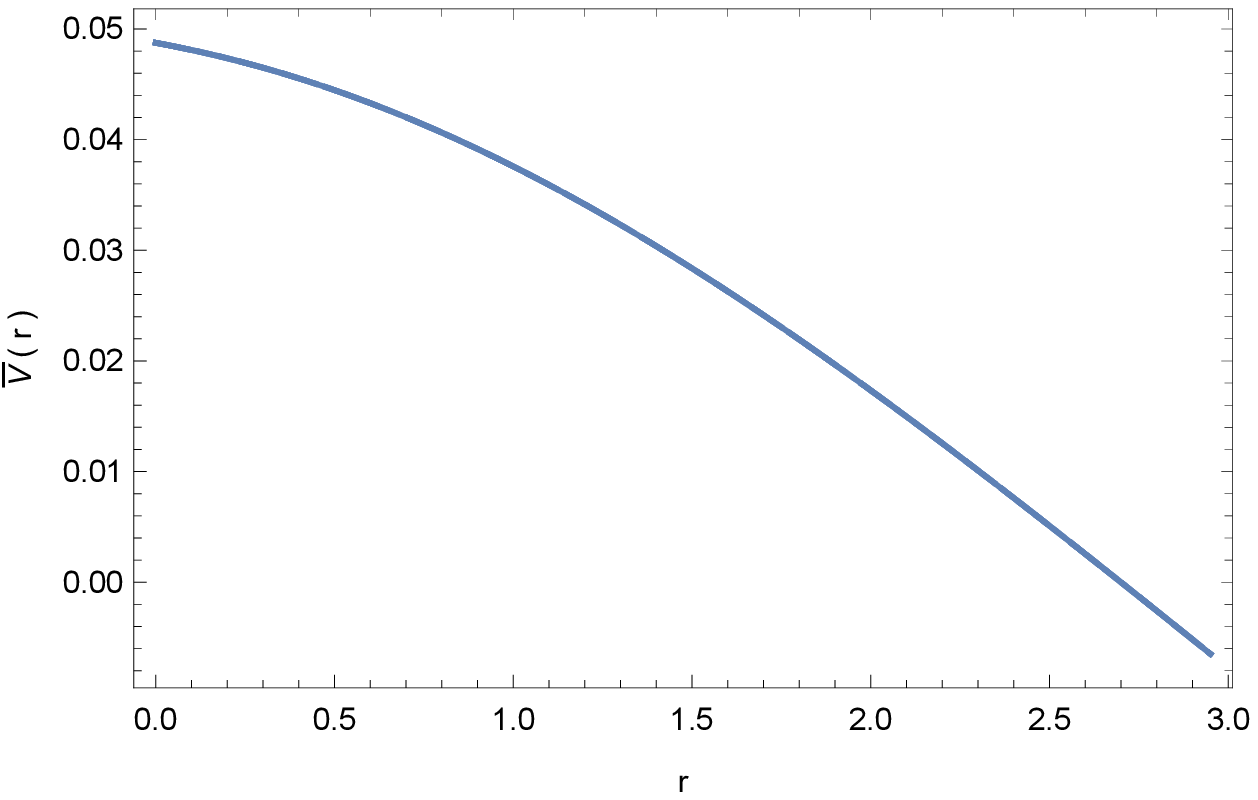}
\includegraphics[width=8.3cm]{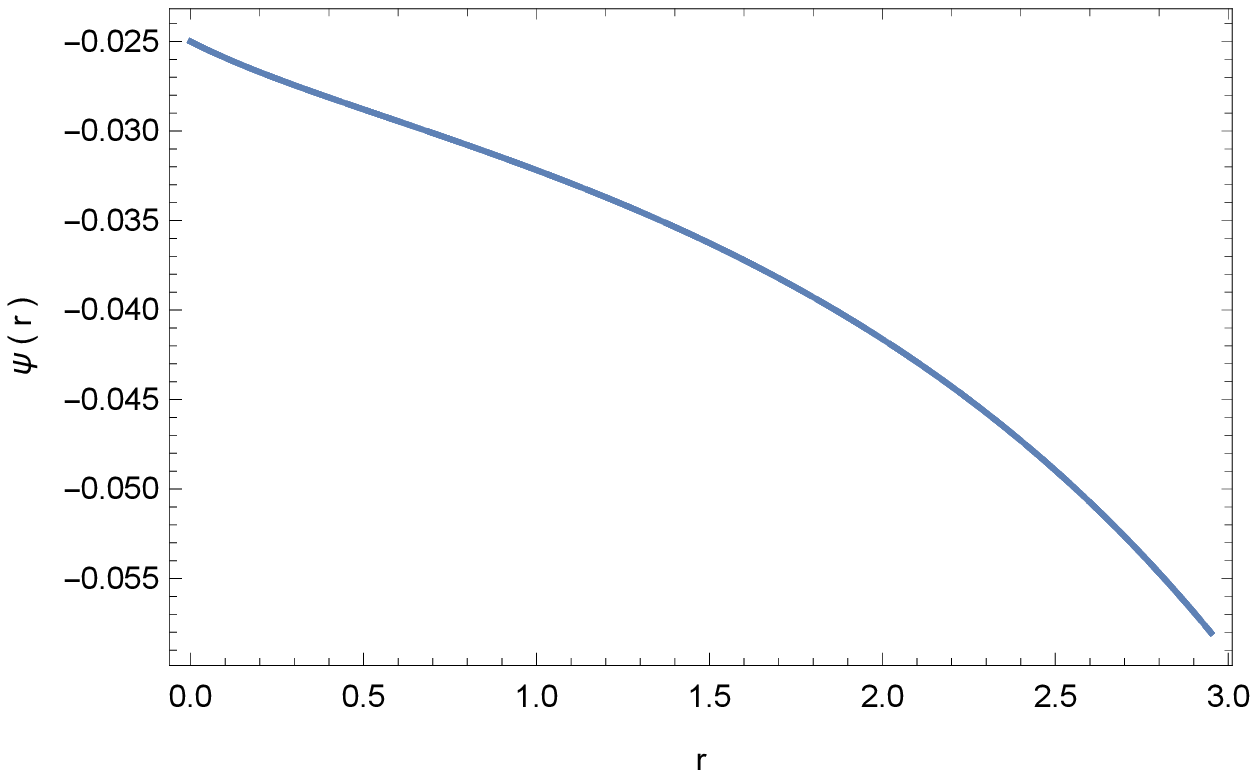}
\caption{Variations of the potential  $\bar{V}\left(\xi,\psi\right)=a\xi ^2+b\psi ^2$ (left panel), and of the function $\psi $ (right panel) as a function of $r$ (with all quantities in arbitrary units)  for $a=0.5$ and $b=-2$ (solid curve), $b=-2.2$ (dotted curve),
 $b=-2.4$ (short dashed curve), $b=-2.6$ (dashed curve), and $b=-2.8$ (long dashed curve), respectively. The boundary conditions used to numerically integrate the field equations are $u_0=-0.01$, $\alpha_0=0.10$, $W(0)=0.15$, and $v_0=-0.01$, respectively. }
\label{fig10}
\end{figure*}

The variations of the potential $\bar{V}$ and of the function $\psi$ are depicted in Fig.~\ref{fig10}. $\bar{V}$ is a decreasing  function of $r$, which near the vacuum boundary of the string takes negative values.   The function $\psi$ is negative inside the string. Both $\bar{V}$ and $\psi$ are basically independent on the variation of the potential parameter $b$.

\begin{figure}[htbp]
\centering
\includegraphics[width=8.3cm]{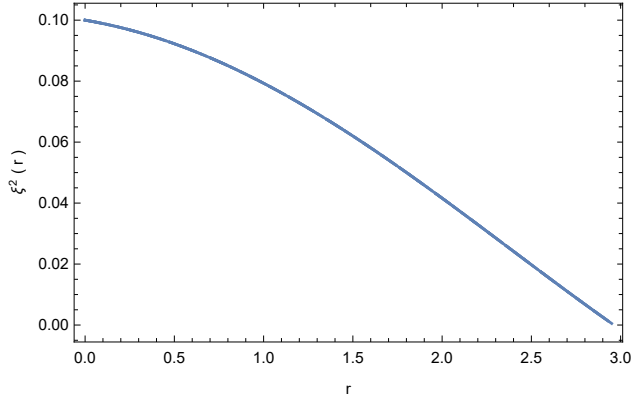}
\caption{Variation of $\xi^2$ as a function of $r$ (with all quantities in arbitrary units) for the $\bar{V}(\xi,\psi)=a\xi^2 +b\psi^2$ potential,  for $a=0.5$ and $b=-2$ (solid curve), $b=-2.2$ (dotted curve),
 $b=-2.4$ (short dashed curve), $b=-2.6$ (dashed curve), and $b=-2.8$ (long dashed curve), respectively. The boundary conditions used to numerically integrate the field equations are $u_0=-0.01$, $\alpha_0=0.10$, $W(0)=0.15$, and $v_0=-0.01$, respectively. }
\label{fig11}
\end{figure}

The behavior of the function $\xi ^2(r)$ is shown in Fig.~\ref{fig11}. $\xi ^2$ is a monotonically decreasing positive function of $r$ for  $r\in \left[0,R_s\right]$, which vanishes on the vacuum boundary of the string $\bar{V}\left(R_s\right)=0$. Inside the string  the gravitational coupling is positive, but in the vacuum outside the string it changes sign. The variation of $\xi^2$ is also basically independent on the numerical values of the potential parameter $b$.

\subsubsection{Varying the initial conditions}

\begin{figure*}[htbp]
\centering
\includegraphics[width=8.3cm]{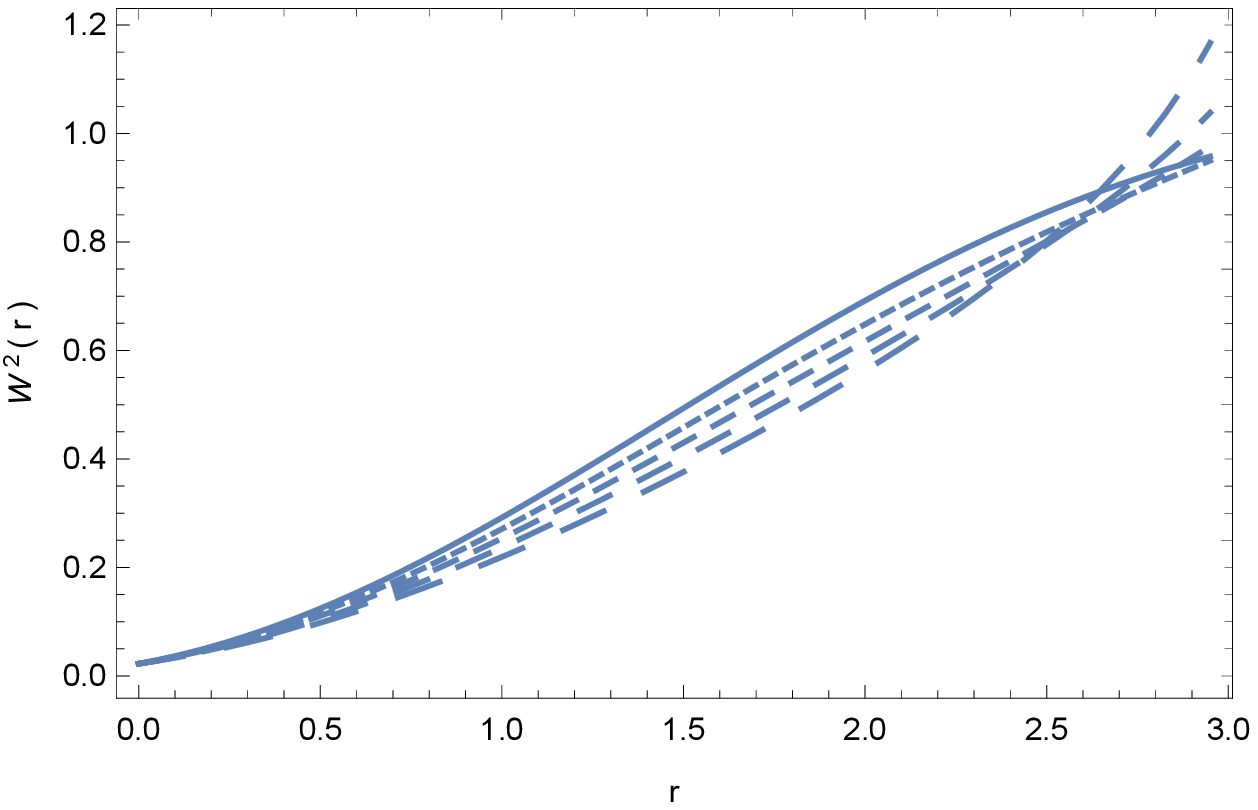}
\includegraphics[width=8.3cm]{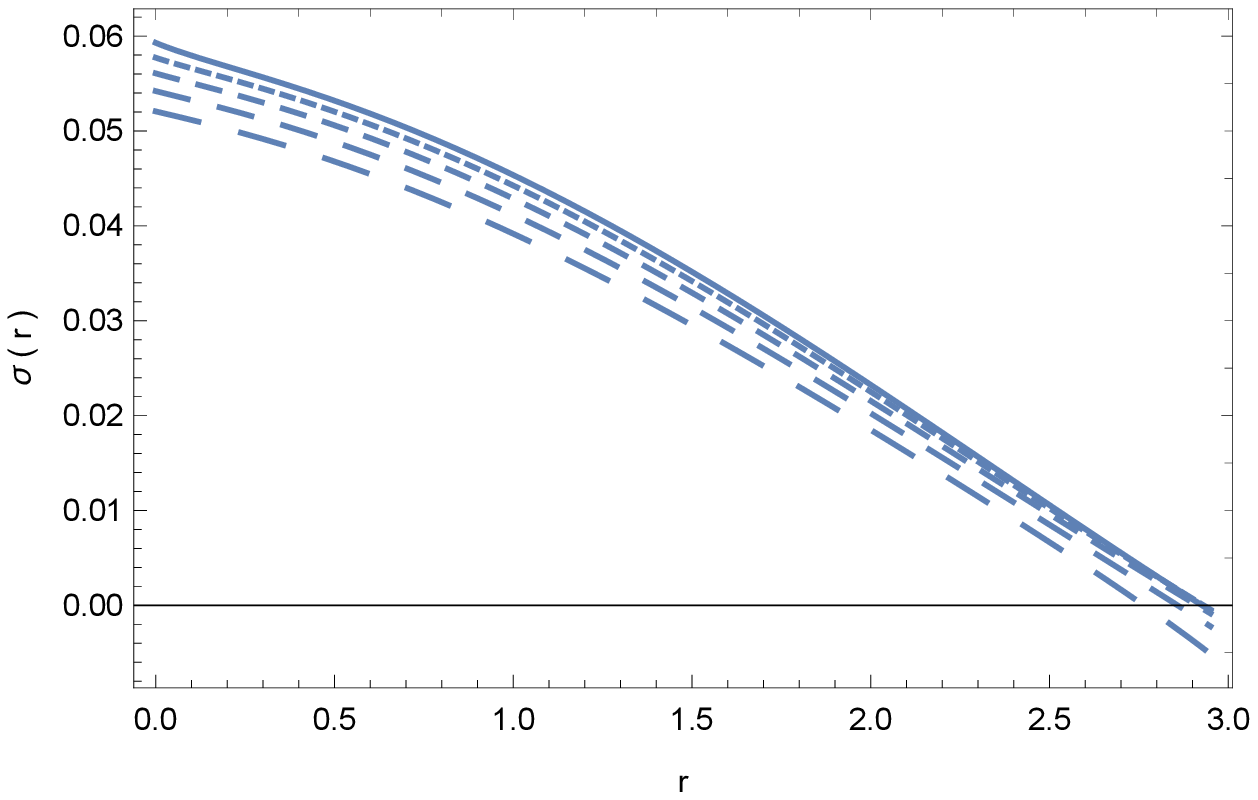}
\caption{Variations of the metric function  $W^2(r)$ (left panel), and of the string tension $\sigma (r) $ (right panel) as a function of $r$ (with all quantities in arbitrary units) for the $\bar{V}(\xi,\psi)=a\xi ^2+b\psi^2$ potential, for $a=0.5$ and $b=-2$, and for different values of $\psi_0$: $\psi_0=-0.025$ (solid curve), $\psi_0=-0.035$ (dotted curve),
$\psi_0=-0.045$ (short dashed curve), $\psi_0=-0.055$ (dashed curve), and $\psi_0=-0.065$ (long dashed curve), respectively. The initial conditions used to numerically integrate the field equations are $u_0=-0.01$, $\alpha_0=0.10$, $W(0)=0.15$, and $v_0=-0.01$, respectively. }
\label{fig12}
\end{figure*}

We consider now the effects of the variation of the initial conditions on the string-like structures in the presence of the quadratic potential $\bar{V}\left(\xi,\psi\right)=a\xi ^2+b\psi ^2$. The variation of the metric function $W^2$ and of the string tension are represented in Fig.~\ref{fig12}, for fixed $a$ and $b$, and varying $\psi '(0)$. The metric function is an increasing function of $r$, and its variation is strongly influenced by the variation of the initial conditions. The string tension is a monotonically decreasing function of $r$ that monotonically decreases, and identically vanishes on the vacuum boundary $r=R_s$ of the string. Thus the string radius is uniquely determined, but its numerical value depends on the adopted initial conditions.

\begin{figure*}[htbp]
\centering
\includegraphics[width=8.3cm]{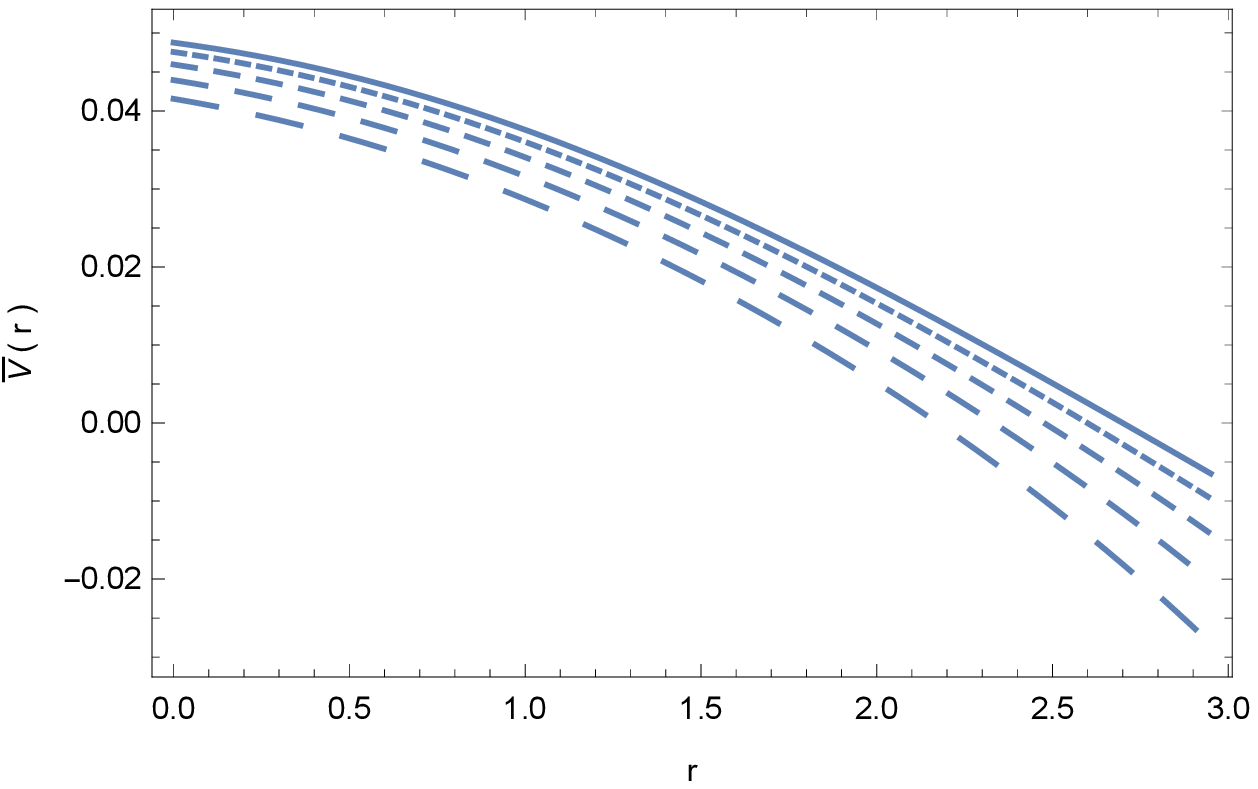}
\includegraphics[width=8.3cm]{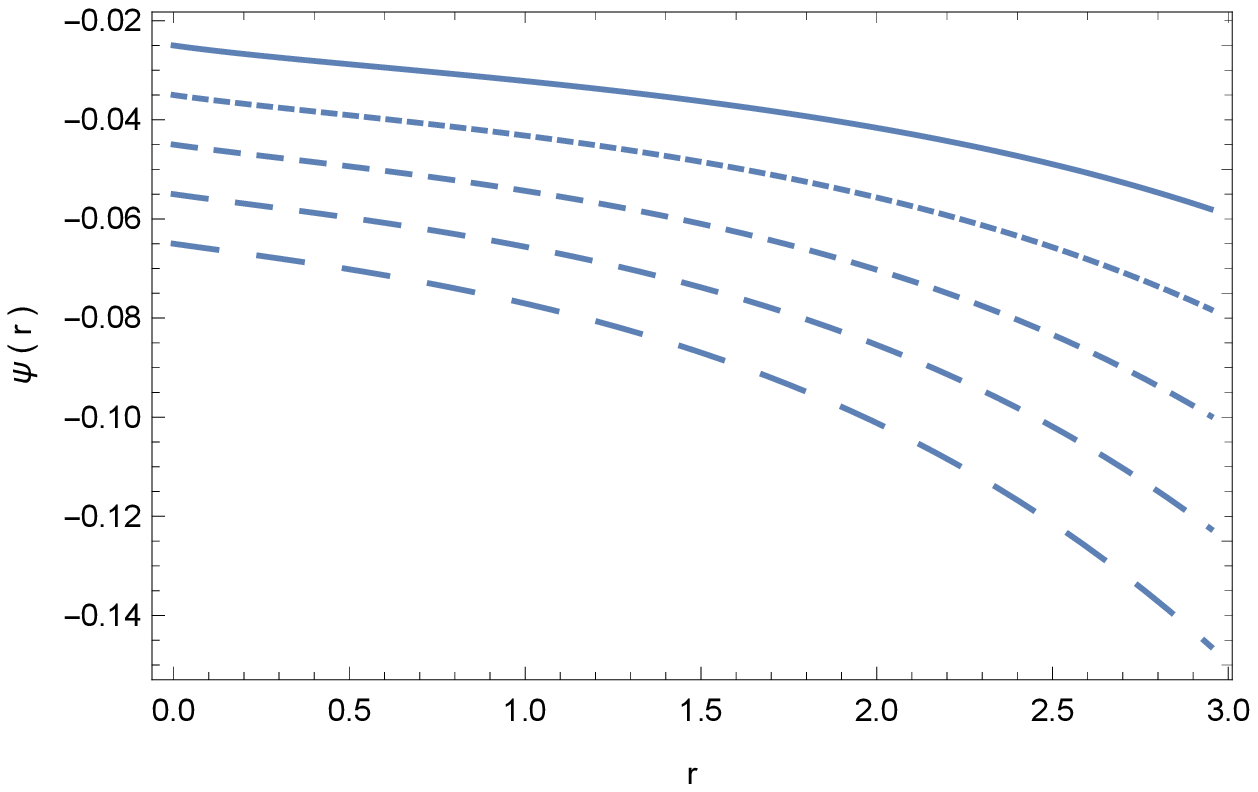}
\caption{Variations of the potential  $\bar{V}\left(\xi,\psi\right)=a\xi ^2+b\psi ^2$ (left panel), and of the function $\psi $ (right panel) as a function of $r$ (with all quantities in arbitrary units) for $a=0.5$ and $b=-2$, and for different values of $\psi_0$: $\psi_0=-0.025$ (solid curve), $\psi_0=-0.035$ (dotted curve),
$\psi_0=-0.045$ (short dashed curve), $\psi_0=-0.055$ (dashed curve), and $\psi_0=-0.065$ (long dashed curve), respectively. The initial conditions used to numerically integrate the field equations are $u_0=-0.01$, $\alpha_0=0.10$, $W(0)=0.15$, and $v_0=-0.01$, respectively. }
\label{fig13}
\end{figure*}

The behaviors of the potential $\bar{V}$ and of the function $\psi$ are shown in Fig.~\ref{fig13}. $\bar{V}$ is a decreasing  function of $r$, which becomes negative near the string boundary.   The function $\psi$ is negative inside the string. Both $\bar{V}$ and $\psi$ are strongly dependent on the numerical value of $\psi _0$.

\begin{figure}[htbp]
\centering
\includegraphics[width=8.3cm]{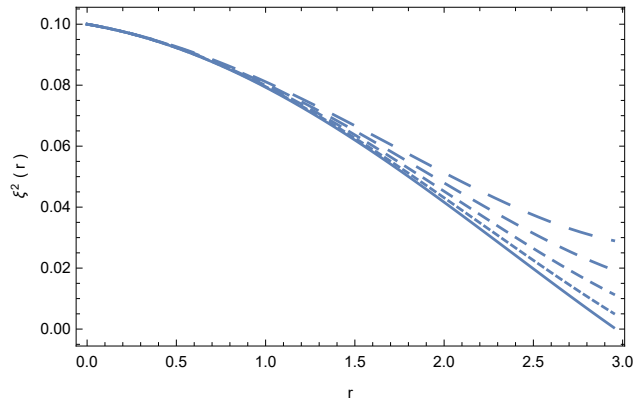}
\caption{Variation of $\xi^2$ as a function of $r$ (with all quantities in arbitrary units) for the $\bar{V}(\xi,\psi)=a\xi^2 +b\psi^2$ potential,  for $a=0.5$ and $b=-2$, and for different values of $\psi_0$: $\psi_0=-0.025$ (solid curve), $\psi_0=-0.035$ (dotted curve),
$\psi_0=-0.045$ (short dashed curve), $\psi_0=-0.055$ (dashed curve), and $\psi_0=-0.065$ (long dashed curve), respectively. The initial conditions used to numerically integrate the field equations are $u_0=-0.01$, $\alpha_0=0.10$, $W(0)=0.15$, and $v_0=-0.01$, respectively. }
\label{fig14}
\end{figure}

The variation of the function $\xi ^2(r)$ with respect to $r$ is represented in Fig.~\ref{fig14}. For all $r$ in the range $r\in\left[0,R_s\right]$, $\xi ^2$ is a monotonically decreasing positive function, which vanishes on the vacuum boundary of the string $\xi^2\left(R_s\right)=0$. Similarly to the previously considered case, inside the string the gravitational coupling is positive. However, for $r>R_s$, $\xi ^2$ changes sign, indicating the presence of the unphysical coupling between the scalar field and the Ricci scalar. The variation of $\xi^2$ is also strongly dependent on the initial conditions.

\section{Discussions and final remarks}\label{sec:conclusion}

Cosmic strings may have been produced in large numbers after the symmetry breaking phase transitions in the postinflationary era. However, the physical and geometrical properties of cosmic strings  may be notably reshaped in modified gravity theories, which have been proposed as alternatives to the $\Lambda$CDM cosmological paradigm. It has already been pointed out that any modifications of canonical GR may strongly affect the gravitational properties of cosmic strings, cylindrically symmetric structures obtained as solutions of the gravitational field equations \cite{Momeni:2015aea, Bailin:1994qt, Goldstone:1962e, Copeland:2009ga, Ringeval:2010ca}.
In the present paper we have considered the properties of cosmic strings in the generalized HMPG theory, a modified gravity theory with the action given by an arbitrary function of the Ricci and Palatini scalars. The equivalent scalar-tensor formulation of the theory is obtained in terms of two scalar fields, as well as of a two-field dependent scalar potential. By assuming a cosmic string metric of the form (\ref{metrn}), obtained by imposing the boost invariance on the general cylindrically symmetric metric, the gravitational field equations can be obtained in terms of the metric tensor component $W^2(r)$, and of the two scalar fields $\varphi$ and $\psi$. However, in the Lagrangian of the scalar-tensor formulation of the theory, the coupling between the scalar fields and the Ricci tensor is of the form $(\phi -\psi)R$. In order to obtain realistic gravitational theories the coupling term $\phi -\psi$ must be positive. This important physical aspect can be implemented by performing a transformation of the fields so that $\phi,\psi)\rightarrow \left(\xi ^2,\psi\right)$, where $\xi ^2=\phi -\psi$. The positivity of the gravitational coupling, giving the physical consistency of the solutions,  can then be formulated as $\xi^2(r)>0, \forall r$. Regions of the space-time where $\xi ^2<0$ can be excluded as unphysical.

The field equations also determine the string tension $\sigma$. These equations must be solved by choosing a functional form for the potential, and by imposing some appropriate boundary conditions at $r=0$ on the two scalar fields $\left(\xi ^2,\psi\right)$, on their derivatives, and for $W^2(0)$. Since in the present approach these boundary conditions are arbitrary, and since the second order, strongly nonlinear system of the gravitational field equations is extremely sensitive to the variation of the boundary conditions, many types of cosmic string structures can be obtained by adopting some specific forms of $V$, and different sets of initial conditions.

The gravitational field equations describing cosmic string structures in generalized HMPG can be solved exactly in the simple cases of the vanishing and constant potentials, respectively. For the case of zero potential, the metric of the string can be written as
\be
ds^2=-dt^2+dr^2+C^2\left(1+\frac{c_2}{r}\right)^2r^2d\theta ^2+dz^2,
\ee
where $c_2$ and $C$ are constants. For $c_2=0$ the metric of the string reduces to the standard general relativistic form, with $C^2=1-\pi G\mu$. However, there is an important difference between the cosmic string solutions in generalized HMPG theory, and standard GR, namely, that the string tension is negative for the solution with $V=0$, and, for $c_2=0$, $-\kappa ^2 \sigma (r)\propto 1/r^2$. In this case the tension diverges for $r=0$. On the other hand, for $c_2\neq 0$, the string tension $\sigma (0)$ is finite (negative), but one needs to impose a boundary value for $W^2(0)$. It has been shown in \cite{Visser:1989kh}  that the energy-momentum tensor of a cubic polyhedral wormhole at the edges of the cube is identical to the energy-momentum tensor of a negative tension classical string. On the other hand, standard field theoretical models of strings predict positive string tensions. However, generalized HMPG theories provide a natural mechanism for generating negative string tension, and hence these models may open some new avenues of research in the field of traversable wormholes.

In the case of the constant, nonzero potential, the field equations can again be solved exactly, and some simple expressions for the geometrical and physical parameters can be obtained. The metric has a similar functional form as for the $V=0$ case, but the string tension can be made positive by an appropriate choice of the potential. A solution with a constant string tension $\kappa ^2\sigma =\Lambda/2$ can also be constructed, as well as a solution having $W(r)=W_0r$, which can describe the standard general relativistic string if $W_0^2=1-8\pi G\mu$.
The string radius $R_s$ can be uniquely defined, and it is given in terms of the constant potential, as well as two integration constants. Under the assumption that the string tension is positive in $r=0$, the string radius is also positive.

For other, more complicated forms of the potential it seems very difficult, if not impossible, to obtain exact analytical solutions of the field equations, and hence they must be solved numerically. In this way it is possible to construct large classes of numerical cosmic string models in generalized HMPG theory. In our study we have considered two distinct types of potentials, having specific algebraic structures in the two scalar fields, and obtained either in an additive, or a multiplicative way. We have considered four such forms of the potential. In the first two cases we have considered that the potential depends on only one of the two scalar fields, so that $\bar{V}(\xi,\psi)=\bar{V}_0\xi^2$, $\bar{V}(\xi,\psi)=\bar{V}_0\xi^4$, and $\bar{V}(\xi,\psi)=\bar{V}_0\psi^2$, respectively. For the other two potentials, for one we have adopted a multiplicative algebraic structure, $\bar{V}(\xi, \psi)=\bar{V}_0\xi ^2\psi ^2$, and an additive structure, $\bar{V}(\xi, \psi)=a\xi ^2+b\psi ^2$, respectively.
 Since the potentials depend on at least one extra constant parameter, together with the five boundary conditions we obtain a very large boundary parameter space, containing from six to nine arbitrary parameters. The large number of parameters allows the construction of a large number of different numerical cosmic string models. However, we have restricted the set of parameters, as well as the physical nature of the solutions, by imposing three physical constraints, namely, that the string tension is positive inside the string, and it vanishes at the vacuum boundary, that the string must have a well defined and unique radius $R_s$, obtained from the condition $\sigma \left(R_s\right)=0$, and that $\xi ^2>0$, $\forall r\in \left[0,R_s\right]$. Even after imposing this set of restrictions, a large variety of string models in generalized HMPG theory can be obtained.

   For the four considered potentials  we have obtained two types of behavior of the metric tensor component $W^2(r)$. It is a monotonically  increasing function of $r$ for the potential $V\left(\xi,\psi\right)-\bar{V}_0\psi ^2$, and a monotonically decreasing function for the other three potentials. For the numerical integration of the field equations we have chosen a nonzero value for $W(0)$, $W(0)\neq 0$, since, at least for the considered cases, no numerical solution satisfying the condition $W(0)=0$ exists (for a more detailed discussion on the existence of an axis or its regularity, refer to \cite{Harko:2020oxq}). From the point of view of the sign of $\sigma (r)$, we have considered only solutions, with $\sigma (r)$ taking positive values, monotonically decreasing inside the string, and vanishing at the vacuum boundary, respectively. The string radius is obtained from the condition $\sigma \left(R_s\right)=0$, which uniquely determines its numerical value.

We have also investigated in detail the effects of the variation of the boundary conditions on the cosmic string configuration. Generally, the string models in generalized HMPG theory are very sensitive to any variations of the boundary conditions for all potentials. The functions $W^2(r)$, $\xi^2 (r)$ and $\psi (r)$ are strongly affected by any small modification of the five boundary values that describe the string configuration. Since the parameter space of the boundary conditions is very large, the results presented in this investigation did not cover all the possible cosmic string structures that could be generated from the variations of the values in the set $\left(\alpha_0,\psi_0, W(0),u_0,v_0 \right)$. Detailed numerical investigations, also involving powerful methods from the mathematical theory of dynamical systems are necessary to give a complete answer to this question.

In conclusion, in the present work we have considered specific cosmic string models that are solutions of the field equations of the generalized HMPG theory. Modified gravity theories may have profound implications on the formation, properties and structure of cosmic strings, interesting and important topological objects that may have been generated in the early Universe.  Hence, the theoretical investigations of strings in modified gravity models may therefore be a worthwhile pathway for future research.

\begin{acknowledgments}
TH is supported by a grant of the Romanian Ministry of Education and Research, CNCS-UEFISCDI, project number PN-III-P4-ID-PCE-2020-2255 (PNCDI III).
FSNL acknowledges support from the Funda\c{c}\~{a}o para a Ci\^{e}ncia e a Tecnologia (FCT) Scientific Employment Stimulus contract with reference CEECINST/00032/2018, and the research grants No. UID/FIS/04434/2020, No. PTDC/FIS-OUT/29048/2017 and No.
CERN/FIS-PAR/0037/2019.  JLR was supported by the European Regional Development Fund and the programme Mobilitas Pluss (MOBJD647)
\end{acknowledgments}




\begin{thebibliography}{99}


\bibitem{Martin:2003cv}
  C. A. Martin, On continuous symmetries and the foundations of modern physics, Symmetries in Physics, edited by
K. Brading and E. Castellani, Cambridge University Press, Cambridge, UK, pp. 29–60.

\bibitem{Weinberg:1967tq}
S.~Weinberg,
``A Model of Leptons,''
Phys. Rev. Lett. \textbf{19}, 1264-1266 (1967).

\bibitem{Salam:1959zz}
A.~Salam and J.~C.~Ward,
``Weak and electromagnetic interactions,''
Nuovo Cim. \textbf{11}, 568-577 (1959).

\bibitem{Salam:1964ry}
A.~Salam and J.~C.~Ward,
``Electromagnetic and weak interactions,''
Phys. Lett. \textbf{13}, 168-171 (1964).

\bibitem{Amaldi:1991zx}
U.~Amaldi, W.~de Boer, P.~H.~Frampton, H.~Furstenau and J.~T.~Liu,
``Consistency checks of grand unified theories,''
Phys. Lett. B \textbf{281}, 374-382 (1992).

\bibitem{Kibble:1976sj}
T.~W.~B.~Kibble,
``Topology of Cosmic Domains and Strings,''
J. Phys. A \textbf{9}, 1387-1398 (1976).

\bibitem{Mermin:1979zz}
N.~D.~Mermin,
``The topological theory of defects in ordered media,''
Rev. Mod. Phys. \textbf{51}, 591-648 (1979).

\bibitem{1975JMoSt..29..190J}
P. G. de Gennes, The Physics of Liquid Crystals, Clarendon Press, Oxford, UK, 1974.

\bibitem{Chuang:1991zz}
I.~Chuang, B.~Yurke, R.~Durrer and N.~Turok,
``Cosmology in the Laboratory: Defect Dynamics in Liquid Crystals,''
Science \textbf{251}, 1336-1342 (1991).

\bibitem{Salomaa:1987zz}
M.~M.~Salomaa and G.~E.~Volovik,
``Quantized vortices in superfluid He-3,''
Rev. Mod. Phys. \textbf{59}, 533-613 (1987)
[erratum: Rev. Mod. Phys. \textbf{60}, 573-573 (1988)].

\bibitem[Hendry et al.(1994)]{1994Natur.368..315H}
P. C. Hendry, N. S. Lawson, R. A. M. Lee, et al.,
``Generation of defects in superfluid $^4$He as an analogue of the formation of cosmic strings,''
Nature, Volume 368, 6469, 315 (1994).

\bibitem{Abrikosov:1956sx}
A.~A.~Abrikosov,
``On the Magnetic properties of superconductors of the second group,''
Sov. Phys. JETP \textbf{5}, 1174-1182 (1957).

\bibitem{Jeannerot:2003qv}
R.~Jeannerot, J.~Rocher and M.~Sakellariadou,
``How generic is cosmic string formation in SUSY GUTs,''
Phys. Rev. D \textbf{68}, 103514 (2003).
[arXiv:hep-ph/0308134 [hep-ph]].

\bibitem{Ade:2013xla}
P.~A.~R.~Ade \textit{et al.} [Planck],
``Planck 2013 results. XXV. Searches for cosmic strings and other topological defects,''
Astron. Astrophys. \textbf{571}, A25 (2014).
[arXiv:1303.5085 [astro-ph.CO]].

\bibitem{Wu:1998mr}
J.~H.~P.~Wu, P.~P.~Avelino, E.~P.~S.~Shellard and B.~Allen,
``Cosmic strings, loops, and linear growth of matter perturbations,''
Int. J. Mod. Phys. D \textbf{11}, 61-102 (2002)
[arXiv:astro-ph/9812156 [astro-ph]].

\bibitem{Olum:2006at}
K.~D.~Olum and A.~Vilenkin,
``Reionization from cosmic string loops,''
Phys. Rev. D \textbf{74}, 063516 (2006)
[arXiv:astro-ph/0605465 [astro-ph]].

\bibitem{Thomas:2009bm}
D.~B.~Thomas, C.~R.~Contaldi and J.~Magueijo,
``Rotation of galaxies as a signature of cosmic strings in weak lensing surveys,''
Phys. Rev. Lett. \textbf{103}, 181301 (2009)
[arXiv:0909.2866 [astro-ph.CO]].

\bibitem{Lake:2015ppa}
M.~J.~Lake and T.~Harko,
``Can Superconducting Cosmic Strings Piercing Seed Black Holes Generate Supermassive Black Holes in the Early Universe?,''
Fortsch. Phys. \textbf{65}, no.10-11, 1600121 (2017)
[arXiv:1505.01584 [astro-ph.CO]].

\bibitem{Binetruy:2012ze}
P.~Binetruy, A.~Bohe, C.~Caprini and J.~F.~Dufaux,
``Cosmological Backgrounds of Gravitational Waves and eLISA/NGO: Phase Transitions, Cosmic Strings and Other Sources,''
JCAP \textbf{06}, 027 (2012)
[arXiv:1201.0983 [gr-qc]].

\bibitem{Weinberg:1974hy}
S.~Weinberg,
``Gauge and Global Symmetries at High Temperature,''
Phys. Rev. D \textbf{9}, 3357-3378 (1974).

\bibitem{Kirzhnits:1974as}
  D.~A.~Kirzhnits and A.~D.~Linde,
  ``A Relativistic phase transition,''
  Sov.\ Phys.\ JETP {\bf 40}, 628 (1975)
  [Zh.\ Eksp.\ Teor.\ Fiz.\  {\bf 67}, 1263 (1974)].

\bibitem{Dolan:1973qd}
L.~Dolan and R.~Jackiw,
``Symmetry Behavior at Finite Temperature,''
Phys. Rev. D \textbf{9}, 3320-3341 (1974).

\bibitem{Vilenkin:2000jqa}
  A.~Vilenkin and E.~P.~S.~Shellard,
  ``Cosmic Strings and Other Topological Defects,''
  Cambridge, UK: Cambridge University Press, 2000., 578.

\bibitem{1991ApJ...373L..35N}
Nambu, Y., Ishihara, H., Gouda, N., et al.,
``Anisotropies of the Cosmic Background Radiation by Domain Wall Networks,''
Astrophysical Journal Letters, 373, L35 (1991).


\bibitem{Vilenkin:1982ks}
A.~Vilenkin and A.~E.~Everett,
``Cosmic Strings and Domain Walls in Models with Goldstone and PseudoGoldstone Bosons,''
Phys. Rev. Lett. \textbf{48}, 1867-1870 (1982).

\bibitem{Nielsen:1973cs}
  H.~B.~Nielsen and P.~Olesen,
  ``Vortex Line Models for Dual Strings,''
  Nucl.\ Phys.\ B {\bf 61}, 45 (1973).

\bibitem{Bucher:1990gs}
  M.~Bucher,
  ``The Aharonov-Bohm effect and exotic statistics for nonabelian vortices,''
  Nucl.\ Phys.\ B {\bf 350}, 163 (1991).

\bibitem{Alford:1992yx}
M.~G.~Alford, K.~M.~Lee, J.~March-Russell and J.~Preskill,
``Quantum field theory of nonAbelian strings and vortices,''
Nucl. Phys. B \textbf{384}, 251-317 (1992)
[arXiv:hep-th/9112038 [hep-th]].

\bibitem{Witten:1984eb}
  E.~Witten,
  ``Superconducting Strings,''
  Nucl.\ Phys.\ B {\bf 249}, 557 (1985).


\bibitem{Christensen:1999wb}
M.~Christensen, A.~L.~Larsen and Y.~Verbin,
``Complete classification of the string - like solutions of the gravitating Abelian Higgs model,''
Phys. Rev. D \textbf{60}, 125012 (1999)
[arXiv:gr-qc/9904049 [gr-qc]].

\bibitem{vandeMeent:2012gb}
M.~van de Meent,
``Geometry of massless cosmic strings,''
Phys. Rev. D \textbf{87}, no.2, 025020 (2013)
[arXiv:1211.4365 [gr-qc]].

\bibitem{Vilenkin:1981zs}
A.~Vilenkin,
``Gravitational Field of Vacuum Domain Walls and Strings,''
Phys. Rev. D \textbf{23}, 852-857 (1981).

\bibitem{Barriola:1989hx}
M.~Barriola and A.~Vilenkin,
``Gravitational Field of a Global Monopole,''
Phys. Rev. Lett. \textbf{63}, 341 (1989).

\bibitem{Sen:1997pu}
A.~A.~Sen, N.~Banerjee and A.~Banerjee,
``Static cosmic strings in Brans-Dicke theory,''
Phys. Rev. D \textbf{56}, 3706-3710 (1997).

\bibitem{Gundlach:1990nm}
C.~Gundlach and M.~E.~Ortiz,
``Jordan-Brans-Dicke cosmic strings,''
Phys. Rev. D \textbf{42}, 2521-2527 (1990).

\bibitem{Barros:1994km}
A.~Barros and C.~Romero,
``Cosmic vacuum strings and domain walls in Brans-Dicke theory of gravity,''
J. Math. Phys. \textbf{36}, 5800-5804 (1995).

\bibitem{Guimaraes:1996ti}
M.~E.~X.~Guimaraes,
``Cosmic string in scalar - tensor gravities,''
Class. Quant. Grav. \textbf{14}, 435-442 (1997)
[arXiv:gr-qc/9610007 [gr-qc]].

\bibitem{Boisseau:1997st}
B.~Boisseau and B.~Linet,
``Dynamics of a selfgravitating thin string in scalar - tensor theories of gravitation,''
Class. Quant. Grav. \textbf{14}, 3063-3071 (1997)
[arXiv:gr-qc/9706020 [gr-qc]].

\bibitem{Dahia:1998vj}
F.~Dahia and C.~Romero,
``Line sources in Brans-Dicke theory of gravity,''
Phys. Rev. D \textbf{60}, 104019 (1999)
[arXiv:gr-qc/9812001 [gr-qc]].

\bibitem{Arazi:2000tn}
A.~Arazi and C.~Simeone,
``Cylindrical sources in full Einstein and Brans-Dicke gravity,''
Gen. Rel. Grav. \textbf{32}, 2259-2268 (2000)
[arXiv:gr-qc/0005026 [gr-qc]].

\bibitem{Gregory:1997wk}
R.~Gregory and C.~Santos,
``Cosmic strings in dilaton gravity,''
Phys. Rev. D \textbf{56}, 1194-1203 (1997)
[arXiv:gr-qc/9701014 [gr-qc]].

\bibitem{Sen:1997hn}
A.~A.~Sen and N.~Banerjee,
``Local cosmic string in generalized scalar tensor theory,''
Phys. Rev. D \textbf{57}, 6558-6560 (1998)
[arXiv:gr-qc/9711036 [gr-qc]].

\bibitem{Sen:1998jj}
A.~A.~Sen,
``Nonstatic local string in Brans-Dicke theory,''
Pramana \textbf{55}, 369-374 (2000)
[arXiv:gr-qc/9803030 [gr-qc]].

\bibitem{Delice:2006gs}
O.~Delice,
``Local cosmic strings in Brans-Dicke theory with cosmological constant,''
Phys. Rev. D \textbf{74}, 067703 (2006)
[arXiv:gr-qc/0609016 [gr-qc]].

\bibitem{Harko:2010mv}
T.~Harko and F.~S.~N.~Lobo,
``f(R,$L_{m}$) gravity,''
Eur. Phys. J. C \textbf{70}, 373-379 (2010)
[arXiv:1008.4193 [gr-qc]].

\bibitem{Harko:2014axa}
T.~Harko and M.~J.~Lake,
``Cosmic strings in $f\left( R,L_m\right) $ gravity,''
Eur. Phys. J. C \textbf{75}, no.2, 60 (2015)
[arXiv:1409.8454 [gr-qc]].

\bibitem{Harko:2020oxq}
T.~Harko, F.~S.~N.~Lobo and H.~M.~R.~da Silva,
``Cosmic stringlike objects in hybrid metric-Palatini gravity,''
Phys. Rev. D \textbf{101}, no.12, 124050 (2020)
[arXiv:2003.09751 [gr-qc]].

\bibitem{Tamanini:2013ltp}
N.~Tamanini and C.~G.~Boehmer,
``Generalized hybrid metric-Palatini gravity,''
Phys. Rev. D \textbf{87}, no.8, 084031 (2013)
[arXiv:1302.2355 [gr-qc]].

\bibitem{Sotiriou:2008rp}
T.~P.~Sotiriou and V.~Faraoni,
``$f(R)$ Theories Of Gravity,''
Rev. Mod. Phys. \textbf{82}, 451-497 (2010)
[arXiv:0805.1726 [gr-qc]].

\bibitem{Khoury:2003aq}
J.~Khoury and A.~Weltman,
``Chameleon fields: Awaiting surprises for tests of gravity in space,''
Phys. Rev. Lett. \textbf{93}, 171104 (2004)
[arXiv:astro-ph/0309300 [astro-ph]].

\bibitem{Khoury:2003rn}
J.~Khoury and A.~Weltman,
``Chameleon cosmology,''
Phys. Rev. D \textbf{69}, 044026 (2004)
[arXiv:astro-ph/0309411 [astro-ph]].

\bibitem{Olmo:2011uz}
G.~J.~Olmo,
``Palatini Approach to Modified Gravity: f(R) Theories and Beyond,''
Int. J. Mod. Phys. D \textbf{20}, 413-462 (2011)
[arXiv:1101.3864 [gr-qc]].

\bibitem{Gomez:2020rnq}
D.~S\'aez-Chill\'on G\'omez,
``Variational principle and boundary terms in gravity à la Palatini,''
Phys. Lett. B \textbf{814}, 136103 (2021)
[arXiv:2011.11568 [gr-qc]].

\bibitem{Harko:2011nh}
T.~Harko, T.~S.~Koivisto, F.~S.~N.~Lobo and G.~J.~Olmo,
``Metric-Palatini gravity unifying local constraints and late-time cosmic acceleration,''
Phys. Rev. D \textbf{85}, 084016 (2012)
[arXiv:1110.1049 [gr-qc]].

\bibitem{Capozziello:2012ny}
S.~Capozziello, T.~Harko, T.~S.~Koivisto, F.~S.~N.~Lobo and G.~J.~Olmo,
``Cosmology of hybrid metric-Palatini f(X)-gravity,''
JCAP \textbf{04}, 011 (2013)
[arXiv:1209.2895 [gr-qc]].

\bibitem{Carloni:2015bua}
S.~Carloni, T.~Koivisto and F.~S.~N.~Lobo,
``Dynamical system analysis of hybrid metric-Palatini cosmologies,''
Phys. Rev. D \textbf{92}, no.6, 064035 (2015)
[arXiv:1507.04306 [gr-qc]].

\bibitem{Capozziello:2012qt}
S.~Capozziello, T.~Harko, T.~S.~Koivisto, F.~S.~N.~Lobo and G.~J.~Olmo,
``The virial theorem and the dark matter problem in hybrid metric-Palatini gravity,''
JCAP \textbf{07}, 024 (2013)
[arXiv:1212.5817 [physics.gen-ph]].

\bibitem{Capozziello:2013yha}
S.~Capozziello, T.~Harko, T.~S.~Koivisto, F.~S.~N.~Lobo and G.~J.~Olmo,
``Galactic rotation curves in hybrid metric-Palatini gravity,''
Astropart. Phys. \textbf{50-52}, 65-75 (2013)
[arXiv:1307.0752 [gr-qc]].

\bibitem{Capozziello:2013uya}
S.~Capozziello, T.~Harko, F.~S.~N.~Lobo and G.~J.~Olmo,
``Hybrid modified gravity unifying local tests, galactic dynamics and late-time cosmic acceleration,''
Int. J. Mod. Phys. D \textbf{22}, 1342006 (2013)
[arXiv:1305.3756 [gr-qc]].

\bibitem{Capozziello:2013wq}
S.~Capozziello, T.~Harko, T.~S.~Koivisto, F.~S.~N.~Lobo and G.~J.~Olmo,
``Hybrid $f(R)$ theories, local constraints, and cosmic speedup,''
[arXiv:1301.2209 [gr-qc]].

\bibitem{Dyadina:2019dsu}
P.~I.~Dyadina, S.~P.~Labazova and S.~O.~Alexeyev,
``Post-Newtonian Limit of Hybrid Metric-Palatini $f(R)$-Gravity,''
J. Exp. Theor. Phys. \textbf{129}, no.5, 838-848 (2019).

\bibitem{Boehmer:2013oxa}
C.~G.~B\"ohmer, F.~S.~N.~Lobo and N.~Tamanini,
``Einstein static Universe in hybrid metric-Palatini gravity,''
Phys. Rev. D \textbf{88}, no.10, 104019 (2013)
[arXiv:1305.0025 [gr-qc]].

\bibitem{Santos:2016tds}
J.~Santos, M.~J.~Rebou\c{c}as and A.~F.~F.~Teixeira,
``Homogeneous G\"odel-type solutions in hybrid metric-Palatini gravity,''
Eur. Phys. J. C \textbf{78}, no.7, 567 (2018)
[arXiv:1611.03985 [gr-qc]].

\bibitem{Kausar:2019iwu}
H.~R.~Kausar, R.~Saleem and A.~Ilyas,
``Cosmological inflation in $f(X)$ gravity theory,''
Phys. Dark Univ. \textbf{26}, 100401 (2019).

\bibitem{Sa:2020qfd}
P.~M.~S\'a,
``Unified description of dark energy and dark matter within the generalized hybrid metric-Palatini theory of gravity,''
Universe \textbf{6}, no.6, 78 (2020)
[arXiv:2002.09446 [gr-qc]].

\bibitem{Sa:2020fvn}
P.~M.~S\'a,
``Triple unification of inflation, dark energy, and dark matter in two-scalar-field cosmology,''
Phys. Rev. D \textbf{102}, no.10, 103519 (2020)
[arXiv:2007.07109 [gr-qc]].

\bibitem{Paliathanasis:2020fyp}
A.~Paliathanasis,
``New cosmological solutions in hybrid metric-Palatini gravity from dynamical symmetries,''
[arXiv:2011.05615 [gr-qc]].

\bibitem{Rosa:2017jld}
J.~L.~Rosa, S.~Carloni, J.~P.~d.~Lemos and F.~S.~N.~Lobo,
``Cosmological solutions in generalized hybrid metric-Palatini gravity,''
Phys. Rev. D \textbf{95}, no.12, 124035 (2017)
[arXiv:1703.03335 [gr-qc]].

\bibitem{Rosa:2019ejh}
J.~L.~Rosa, S.~Carloni and J.~P.~S.~Lemos,
``Cosmological phase space of generalized hybrid metric-Palatini theories of gravity,''
Phys. Rev. D \textbf{101}, no.10, 104056 (2020)
[arXiv:1908.07778 [gr-qc]].

\bibitem{Rosa:2021ish}
J.~L.~Rosa, F.~S.~N.~Lobo and D.~Rubiera-Garcia,
``Sudden singularities in generalized hybrid metric-Palatini cosmologies,''
[arXiv:2103.02580 [gr-qc]].

\bibitem{Lima:2014aza}
N.~A.~Lima,
``Dynamics of Linear Perturbations in the hybrid metric-Palatini gravity,''
Phys. Rev. D \textbf{89}, no.8, 083527 (2014)
[arXiv:1402.4458 [astro-ph.CO]].

\bibitem{Lima:2015nma}
N.~A.~Lima and V.~S.-Barreto,
``Constraints on Hybrid Metric-palatini Gravity from Background Evolution,''
Astrophys. J. \textbf{818}, no.2, 186 (2016)
[arXiv:1501.05786 [astro-ph.CO]].

\bibitem{Leanizbarrutia:2017xyd}
I.~Leanizbarrutia, F.~S.~N.~Lobo and D.~Saez-Gomez,
``Crossing SNe Ia and BAO observational constraints with local ones in hybrid metric-Palatini gravity,''
Phys. Rev. D \textbf{95}, no.8, 084046 (2017)
[arXiv:1701.08980 [gr-qc]].

\bibitem{Avdeev:2020jqo}
N.~Avdeev, P.~Dyadina and S.~Labazova,
``Test of hybrid metric-Palatini f(R)-gravity in binary pulsars,''
J. Exp. Theor. Phys. \textbf{131}, no.4, 537-547 (2020)
[arXiv:2009.11156 [gr-qc]].

\bibitem{Koivisto:2013kwa}
T.~S.~Koivisto and N.~Tamanini,
``Ghosts in pure and hybrid formalisms of gravity theories: A unified analysis,''
Phys. Rev. D \textbf{87}, no.10, 104030 (2013)
[arXiv:1304.3607 [gr-qc]].

\bibitem{Capozziello:2013gza}
S.~Capozziello, T.~Harko, F.~S.~N.~Lobo, G.~J.~Olmo and S.~Vignolo,
``The Cauchy problem in hybrid metric-Palatini f(X)-gravity,''
Int. J. Geom. Meth. Mod. Phys. \textbf{11}, no.5, 1450042 (2014)
[arXiv:1312.1320 [gr-qc]].

\bibitem{Chen:2020evr}
C.~Y.~Chen, Y.~H.~Kung and P.~Chen,
``Black Hole Perturbations and Quasinormal Modes in Hybrid Metric-Palatini Gravity,''
Phys. Rev. D \textbf{102}, no.12, 124033 (2020)
[arXiv:2010.07202 [gr-qc]].

\bibitem{Danila:2018xya}
B.~Dǎnilǎ, T.~Harko, F.~S.~N.~Lobo and M.~K.~Mak,
``Spherically symmetric static vacuum solutions in hybrid metric-Palatini gravity,''
Phys. Rev. D \textbf{99}, no.6, 064028 (2019)
[arXiv:1811.02742 [gr-qc]].

\bibitem{Bronnikov:2019ugl}
K.~A.~Bronnikov,
``Spherically symmetric black holes and wormholes in hybrid metric-Palatini gravity,''
Grav. Cosmol. \textbf{25}, 331-341 (2019)
[arXiv:1908.02012 [gr-qc]].

\bibitem{Bronnikov:2020vgg}
K.~A.~Bronnikov, S.~V.~Bolokhov and M.~V.~Skvortsova,
``Hybrid metric-Palatini gravity: black holes, wormholes, singularities and instabilities,''
Grav. Cosmol. \textbf{26}, no.3, 212-227 (2020)
[arXiv:2006.00559 [gr-qc]].

\bibitem{Rosa:2020uoi}
J.~L.~Rosa, J.~P.~S.~Lemos and F.~S.~N.~Lobo,
``Stability of Kerr black holes in generalized hybrid metric-Palatini gravity,''
Phys. Rev. D \textbf{101}, 044055 (2020)
[arXiv:2003.00090 [gr-qc]].

\bibitem{Danila:2016lqx}
B.~Danila, T.~Harko, F.~S.~N.~Lobo and M.~K.~Mak,
``Hybrid metric-Palatini stars,''
Phys. Rev. D \textbf{95}, no.4, 044031 (2017)
[arXiv:1608.02783 [gr-qc]].

\bibitem{Capozziello:2012hr}
S.~Capozziello, T.~Harko, T.~S.~Koivisto, F.~S.~N.~Lobo and G.~J.~Olmo,
``Wormholes supported by hybrid metric-Palatini gravity,''
Phys. Rev. D \textbf{86}, 127504 (2012)
[arXiv:1209.5862 [gr-qc]].

\bibitem{KordZangeneh:2020ixt}
M.~Kord Zangeneh and F.~S.~N.~Lobo,
``Dynamic wormhole geometries in hybrid metric-Palatini gravity,''
Eur. Phys. J. C \textbf{81}, no.4, 285 (2021)
[arXiv:2011.01745 [gr-qc]].

\bibitem{Rosa:2018jwp}
J.~L.~Rosa, J.~P.~S.~Lemos and F.~S.~N.~Lobo,
``Wormholes in generalized hybrid metric-Palatini gravity obeying the matter null energy condition everywhere,''
Phys. Rev. D \textbf{98}, no.6, 064054 (2018)
[arXiv:1808.08975 [gr-qc]].

\bibitem{Rosa:2021yym}
J.~L.~Rosa,
Phys. Rev. D \textbf{104} (2021) no.6, 064002
doi:10.1103/PhysRevD.104.064002
[arXiv:2107.14225 [gr-qc]].

\bibitem{Rosa:2021lhc}
J.~L.~Rosa, F.~S.~N.~Lobo and G.~J.~Olmo,
``Weak-field regime of the generalized hybrid metric-Palatini gravity,''
[arXiv:2104.10890 [gr-qc]].

\bibitem{Kausar:2018ipo}
H.~R.~Kausar,
``Gravitational wave solutions in hybrid metric-Palatini theory,''
Astrophys. Space Sci. \textbf{363}, no.11, 238 (2018).

\bibitem{Bombacigno:2019did}
F.~Bombacigno, F.~Moretti and G.~Montani,
``Scalar modes in extended hybrid metric-Palatini gravity: weak field phenomenology,''
Phys. Rev. D \textbf{100}, no.12, 124036 (2019)
[arXiv:1907.11949 [gr-qc]].

\bibitem{Fu:2016szo}
Q.~M.~Fu, L.~Zhao, B.~M.~Gu, K.~Yang and Y.~X.~Liu,
``Hybrid metric-Palatini brane system,''
Phys. Rev. D \textbf{94}, no.2, 024020 (2016)
[arXiv:1601.06546 [gr-qc]].

\bibitem{Rosa:2020uli}
J.~L.~Rosa, D.~A.~Ferreira, D.~Bazeia and F.~S.~N.~Lobo,
``Thick brane structures in generalized hybrid metric-Palatini gravity,''
Eur. Phys. J. C \textbf{81}, no.1, 20 (2021)
[arXiv:2010.10074 [gr-qc]].

\bibitem{Harko:2018ayt}
    T.~Harko, F.~S.~N.~Lobo. {\it Extensions of $f(R)$ Gravity:
    Curvature-Matter Couplings and Hybrid Metric-Palatini Gravity}
    (Cambridge Monographs on Mathematical Physics). Cambridge:
    Cambridge University Press (2018).

\bibitem{Capozziello:2015lza}
S.~Capozziello, T.~Harko, T.~S.~Koivisto, F.~S.~N.~Lobo and G.~J.~Olmo,
``Hybrid metric-Palatini gravity,''
Universe \textbf{1}, no.2, 199-238 (2015)
[arXiv:1508.04641 [gr-qc]].

\bibitem{Harko:2020ibn}
T.~Harko and F.~S.~N.~Lobo,
``Beyond Einstein\textquoteright{}s General Relativity: Hybrid metric-Palatini gravity and curvature-matter couplings,''
Int. J. Mod. Phys. D \textbf{29} (2020) no.13, 2030008
[arXiv:2007.15345 [gr-qc]].



\bibitem{Abbott:2017mem}
B.~P.~Abbott \textit{et al.} [LIGO Scientific and Virgo],
``Constraints on cosmic strings using data from the first Advanced LIGO observing run,''
Phys. Rev. D \textbf{97}, no.10, 102002 (2018)
[arXiv:1712.01168 [gr-qc]].

\bibitem{Battye:2010xz}
R.~Battye and A.~Moss,
``Updated constraints on the cosmic string tension,''
Phys. Rev. D \textbf{82}, 023521 (2010)
[arXiv:1005.0479 [astro-ph.CO]].

\bibitem{Mack:2007ae}
K.~J.~Mack, D.~H.~Wesley and L.~J.~King,
``Observing cosmic string loops with gravitational lensing surveys,''
Phys. Rev. D \textbf{76}, 123515 (2007)
[arXiv:astro-ph/0702648 [astro-ph]].

\bibitem{Khatri:2008zw}
R.~Khatri and B.~D.~Wandelt,
``Cosmic (super)string constraints from 21 cm radiation,''
Phys. Rev. Lett. \textbf{100}, 091302 (2008)
[arXiv:0801.4406 [astro-ph]].

\bibitem{Caprini:2015zlo}
C.~Caprini, M.~Hindmarsh, S.~Huber, T.~Konstandin, J.~Kozaczuk, G.~Nardini, J.~M.~No, A.~Petiteau, P.~Schwaller and G.~Servant, \textit{et al.}
``Science with the space-based interferometer eLISA. II: Gravitational waves from cosmological phase transitions,''
JCAP \textbf{04}, 001 (2016)
[arXiv:1512.06239 [astro-ph.CO]].

\bibitem{Visser:1989kh}
M.~Visser,
``Traversable wormholes: Some simple examples,''
Phys. Rev. D \textbf{39}, 3182-3184 (1989)
[arXiv:0809.0907 [gr-qc]].

\bibitem{Cramer:1994qj}
J.~G.~Cramer, R.~L.~Forward, M.~S.~Morris, M.~Visser, G.~Benford and G.~A.~Landis,
``Natural wormholes as gravitational lenses,''
Phys. Rev. D \textbf{51}, 3117-3120 (1995)
[arXiv:astro-ph/9409051 [astro-ph]].

\bibitem{Momeni:2015aea}
D.~Momeni, K.~Myrzakulov, R.~Myrzakulov and M.~Raza,
``Cylindrical solutions in Mimetic gravity,''
Eur. Phys. J. C \textbf{76}, no.6, 301 (2016)
[arXiv:1505.08034 [gr-qc]].

\bibitem{Bailin:1994qt}
  D.~Bailin and A.~Love,
  ``Supersymmetric gauge field theory and string theory,''
  Bristol, UK: Institute of Physics Pub. IOP (1994) 322 p. (Graduate student series in physics).
  
\bibitem{Goldstone:1962es}
J.~Goldstone, A.~Salam and S.~Weinberg,
Phys. Rev. \textbf{127}, 965-970 (1962).

\bibitem{Copeland:2009ga}
E.~J.~Copeland and T.~W.~B.~Kibble,
``Cosmic Strings and Superstrings,''
Proc. Roy. Soc. Lond. A \textbf{466}, 623-657 (2010)
[arXiv:0911.1345 [hep-th]].

\bibitem{Ringeval:2010ca}
C.~Ringeval,
``Cosmic strings and their induced non-Gaussianities in the cosmic microwave background,''
Adv. Astron. \textbf{2010}, 380507 (2010)
[arXiv:1005.4842 [astro-ph.CO]].

\end{thebibliography}
\end{document}